\documentclass[twocolumn,showpacs,preprintnumbers,
pra,amsmath,amssymb,superscriptaddress]{revtex4}

\usepackage{bm}
\usepackage{epsfig}

\newcommand{\bfsfD}{\mbox{\sffamily\bfseries{D}}}
\newcommand{\bfsfG}{\mbox{\sffamily\bfseries{G}}}
\newcommand{\bfsfI}{\mbox{\sffamily\bfseries{I}}}

\newcommand{\bfsfL}{\mbox{\sffamily\bfseries{L}}}
\newcommand{\bfsfT}{\mbox{\sffamily\bfseries{T}}}
\newcommand{\bfsfV}{\mbox{\sffamily\bfseries{V}}}
\newcommand{\bfmu}{\bm \mu}
\newcommand{\bfsigma}{\bm \sigma}
\newcommand{\bftau}{\bm \tau}

\begin{document}
\title{Spontaneous-emission rates
 in finite photonic crystals of plane scatterers}
\date{Submitted to Phys. Rev. E on 1 July 2003}
\author{Martijn Wubs} \email{c.m.wubs@tn.utwente.nl}
\homepage{http://tnweb.tn.utwente.nl/cops/} \affiliation{Complex
Photonic Systems, Faculty of Science and Technology, University of
Twente, P.O. Box 217, NL-7500~AE~~Enschede, The Netherlands}
\affiliation{Van der Waals-Zeeman Institute, University of
Amsterdam, Valckenierstraat 65, NL-1018 XE  Amsterdam, The
Netherlands}
\author{L.G.~Suttorp}
\affiliation{Institute for Theoretical Physics, University of
Amsterdam, Valckenierstraat 65, NL-1018 XE  Amsterdam, The
Netherlands}
\author{A.~Lagendijk}
\affiliation{Complex Photonic Systems, Faculty of Science and
Technology, University of Twente, P.O. Box 217,
NL-7500~AE~~Enschede, The Netherlands}

\begin{abstract}

 The concept of a plane scatterer that was developed earlier
 for scalar waves is generalized so that polarization of light is included.
Starting from a Lippmann-Schwinger formalism for vector waves, we
show that the Green function has to be regularized before
T-matrices can be defined in a consistent way. After the
regularization, optical modes and Green functions are determined
exactly for finite structures built up of an arbitrary number of
parallel planes, at arbitrary positions, and where each plane can
have different optical properties. The model is applied to the
special case of finite crystals consisting of regularly spaced
 identical planes, where analytical methods can be taken
further and only light numerical tasks remain.  The formalism is
used to calculate position- and orientation-dependent
spontaneous-emission rates inside and near the finite photonic
crystals. The results show that emission rates and reflection
properties can differ strongly for scalar and for vector waves.
The finite size of the crystal influences the emission rates. For
parallel dipoles close to a plane, emission into guided modes
gives rise to a peak in the frequency-dependent emission rate.
\end{abstract}


\maketitle

\section{Introduction}\label{Introduction}
Photonic crystals are a well-studied subject nowadays, both
theoretically and experimentally \cite{Soukoulis01}. Of
fundamental importance is the prediction \cite{YabJohn87} that in
three-dimensional photonic crystals that meet a tough combination
of requirements, light propagation will be completely inhibited in
all  directions and a photonic band gap will show up for certain
frequencies of light. It is important for technology  that
photonic crystals can be created that  guide light with low
losses, and bend light on a
 scale of an optical wavelength. The latter properties do not
require a  band gap in all three dimensions.

 A photonic-band-gap crystal would reflect light for all angles of incidence, when the
  frequency of the light lies within the gap. However, lower-dimensional photonic
 crystals such as  Bragg mirrors can also be omnidirectional
 mirrors, without having a band gap \cite{Fink98,Hooijer00,Wubs02}.
 Thus, external light sources can only give an
 indication that there is a band gap or a proof that there is no gap.

Internal light sources such as excited atoms do a better job in
probing a  band
 gap, because only a gap would completely inhibit spontaneous emission
by internal sources \cite{YabJohn87}. For the same reason, a
photonic-band-gap crystal would be a whole new playground in
quantum optics, both when one is interested in spontaneous
emission in itself, and in processes which normally are obscured
or made less efficient because of spontaneous emission. Not only
emission rates would be strongly modified inside a band-gap
crystal,  but also resonant dipole-dipole interactions, for
example, as they are mediated by the electromagnetic field
\cite{dipdipref}. The focus of this paper is on
spontaneous-emission rates of visible light.

For atomic transition frequencies in the band gap of an infinite
three-dimensional photonic
 crystal, emission rates vanish everywhere in the
inhomogeneous structure. In practice, such a uniform suppression
of emission rates has not yet been observed for visible light:
evidence of crystals exhibiting a  full photonic band gap in the
visible has not been reported to date. Even when in the  future
such crystals will exist, position-dependent emission rates will
occur at the edges of the crystals. In general,
spontaneous-emission rates of inhomogeneous dielectrics with a
high refractive-index contrast, including photonic crystals, are
strongly position and orientation dependent. Calculated
spontaneous-emission rates in this paper will prove this point.
Also, finite-size effects will show up in our calculations. The
model studied here is a finite photonic crystal consisting of a
finite number of parallel and infinitely thin planes. More about
our model will be said later in this Introduction.

In many experiments, dipole orientations are hard to control. When
averaged over dipole orientations, spontaneous-emission rates are
proportional to a quantity called the `local optical density of
states' (LDOS) \cite{Sprik96,Busch98}. The concept of a local
density of states was borrowed from solid-state physics. The local
optical density of states was first named the `local radiative
density of states' \cite{Sprik96}, which is the same quantity.

Interestingly, the calculation of position-dependent
spontaneous-emission rates also has a bearing on the
interpretation of measurements performed with a near-field
scanning optical microscope (or NSOM). In these measurements, a
sample is illuminated through the tip of the microscope, and
scattered light is recorded. In a simple model, the disturbance of
the optical field by bringing the tip of the microscope to the
sample is assumed to be weak, and the tip  is modelled as a dipole
with a certain strength and orientation. Then, if  the light
scattered in all directions would be recorded, the measured signal
would be proportional to the local spontaneous-emission rate at
the position of the tip in the absence of the tip \cite{Colas01}.

After the above general considerations, we now turn to the topic
how spontaneous-emission rates inside  photonic crystals are
actually being calculated. The existence of a band gap in an
infinite photonic crystal can be inferred from a band structure
calculation, which for a three-dimensional photonic crystals is an
art in itself (see the recent review \cite{Busch02}). Quite
another and more difficult matter is it to calculate  emission
rates
 inside infinite crystals \cite{Suzuki95,Li00_01}. Emission rates
inside or near {\em finite} photonic crystals are even harder to
calculate. Other interesting quantities  would be near-field or
far-field spectra of internal sources, or dipole-dipole
interactions and superradiance effects of
 atoms embedded in a finite three-dimensional photonic crystal,
 to name a few complex processes
in a complex environment. In such cases, results of calculations
are hard to check and - even if correct - they might not give much
insight.

It is therefore very useful to study  complex processes  in
simplified models for photonic crystals. Widely used is the
so-called quasi one-dimensional model (or isotropic model) for
photonic crystals \cite{quasionedimensional}, where it is assumed
that the red edge of the stop bands of the crystal occur at the
same band-edge frequency for all three-dimensional propagation
directions, and similarly for the blue band edge. Such a model
will describe qualitatively correct the processes well inside the
band gap, while overestimating effects of the photonic crystal  at
the edges of the  gap. The isotropic model also neglects all
position and orientation dependence of emission rates outside the
band gap. Inspired by the model calculations, more realistic
numerical calculations have recently appeared that indeed show the
weaknesses of the isotropic model \cite{Li00_01}.

In this paper another simple model is proposed, one which takes
into account the strong spatial and orientational dependence of
optical properties and the finite size of the  crystals. On the
other hand, it gives up the existence of a full band gap, as only
variations of the refractive index in one dimension are
considered. Dielectric slabs are modelled as infinitely thin
planes, which will be called {\em  plane scatterers}.  A
multiple-scattering formalism is set up in which optical modes and
the Green function (a tensor, really) can be calculated exactly
for crystals consisting of an arbitrary number of plane
scatterers. The present model is a generalization of previous work
that treated scalar waves only \cite{Wubs02}. The inclusion of
polarization of light  will turn out not to be straightforward.

Infinitely thin planes were used as model systems in photonics
before, for example  in \cite{Dowling92} where light propagation
was considered  in one dimension only and for an infinite crystal.
The model was generalized in  \cite{Shepherd97}, where infinite
photonic crystals were built of infinitely thin planes and their
band structure was determined for waves propagating in three
dimensions. In \cite{Visser97}, both infinite and finite crystals
of planes were considered and their transmission and reflection
properties were studied with the use of transfer matrix methods.
The infinite crystal was again considered in
\cite{Alvarado99,Zurita00,Alvarado01,Zurita02} and named the
`Dirac-comb superlattice'. Frequency-dependent emission rates were
determined for several positions in the unit cell and both TE-
\cite{Alvarado99,Alvarado01} and TM-waves \cite{Zurita00,Zurita02}
were considered. The periodicity of infinite crystals makes that
Bloch's theorem can be used in the analysis. Finite photonic
crystals do not have this advantage and the analysis of the
optical properties is usually more difficult.  Position-dependent
spontaneous-emission rates remain to be explored in model photonic
crystals consisting of a finite number of plane scatterers, where
light propagation in all three dimensions is taken into account,
for both polarization directions.

It has been known for a long time that spontaneous-emission rates
of atoms change when positioned at distances on the order of the
wavelength of light away from a mirror
\cite{Drexhage70,Haroche92,Milonni94}. A recent surprise was the
 measurement and analysis that even a distant mirror ($25 {\rm
 cm}$
away) can change emission rates when lenses are used and when the
atomic positions are controlled with a sub-wavelength precision
(a few nanometers) \cite{Eschner01,Beige02}.

More complicated than emission near a mirror is it to calculate
emission rates of atoms in or near one-dimensional photonic
crystals. A multi-purpose formalism for calculating optical modes
in layered dielectrics \cite{Tomas95} was used in \cite{Hooijer00}
to calculate emission rates inside  finite periodic layered
structures, especially inside structures that reflect light
incoming from all directions, the so-called `omnidirectional
mirrors' \cite{Fink98}.

Interestingly, light transmission through finite one-dimensional
photonic crystals can be found exactly in terms of the
transmission through a unit cell, the number $N$ of unit cells,
and in terms of the Bloch wave vector of the corresponding
infinite crystal structure. In \cite{Yariv84}, this is shown for a
simple unit cell containing two layers, but it was also proven for
general  unit cells \cite{Bendickson96}. This remarkable result
was reviewed in \cite{Griffiths00}, where its importance is
stressed not only in optics but also in acoustics, quantum
mechanics, and other branches of physics. In the T-matrix
formalism of this paper (which differs from the usual transfer
matrix method for layered dielectrics), we find similar analytical
results, also involving the Bloch wave vector. Such attractive
analytical results are not available for  more complex dielectrics
such as finite two- \cite{Asatryan,Martin99} or three-dimensional
photonic crystals \cite{Doosje02}, so that in those cases the use
of efficient numerical techniques is essential.

The advantage of our plane-scatterer model is that modes and Green
functions (and therefore emission rates) can be calculated exactly
in a Lippmann-Schwinger formalism, for every finite crystal size,
and that light propagation  in all directions is taken into
account.  Lippmann-Schwinger formalisms are more commonly used
\cite{Gonis00}, but when the finite volumes of dielectric
scatterers are fully taken into account,  numerical discretization
of the dielectric is required and the model stops being simple
\cite{Martin99,Sondergaard02}. To be sure, the simplicity of our
model entails that in some aspects it becomes less realistic, as
will be stressed where appropriate.

In Sec.~\ref{multiplevector}, multiple scattering of light is
introduced and central equations are derived in
representation-independent notation. In Sec.~\ref{planescatforvec}
the free-space Green tensor is regularized and a T-matrix of a
plane scatterer for light waves is derived. Sec.~\ref{modesvector}
discusses all optical modes (propagating and guided modes,
including polarization) that exist in crystals of plane
scatterers. Position and orientation dependent
spontaneous-emission rates are calculated in Sec.~\ref{SEvector}.
Conclusions can be found in Sec.~\ref{concchapvecplanes}.

\section{Multiple-scattering theory for vector
waves}\label{multiplevector} Some important equations of
multiple-scattering theory \cite{Gonis00} will be presented,
mostly in representation-independent notation, for light in
arbitrary inhomogeneous dielectrics. In later sections a
particular class of dielectrics will be studied and a suitable
representation is chosen, but then the involved notation might
obscure the general structure of the equations.

 The wave equation for the
electric field ${\bf E}_{0}({\bf r},\omega)$ in free space is

\begin{subequations}\begin{eqnarray}\label{efieldvac} (\omega/c)^{2}{\bf
E}_{0}({\bf r},\omega)-\nabla \times \nabla
\times {\bf E}_{0}({\bf r},\omega) = 0 \Leftrightarrow \label{efieldvac_a} \\
 \bigl\{  (\omega/c)^{2} \bfsfI \cdot - \nabla \times \nabla
\times \bigl\} {\bf E}_{0}({\bf r},\omega) = 0.
\label{efieldvac_b}
\end{eqnarray}
\end{subequations}
The symbol $\bfsfI$ denotes the unit tensor in three-dimensional
space. The solutions of Eq.~(\ref{efieldvac}) are plane waves with
wave vector ${\bf k}$ and polarization direction normal to ${\bf
k}$. With the free-space wave equation~(\ref{efieldvac}), a Green
tensor (or dyadic Green function)  is associated that satisfies
\begin{equation}\label{G0free}
\bigl\{  (\omega/c)^{2} \bfsfI \cdot - \nabla \times \nabla \times
\bigl\}\bfsfG_{0}({\bf r},{\bf r'},\omega) = \delta^{3}({\bf
r}-{\bf r'})\bfsfI.
\end{equation}
Let  $\bfsfL({\bf r},\omega)$ be the  quantity between curly
brackets in Eqs.~(\ref{efieldvac_b}) and (\ref{G0free}). Both
equations can be considered as the real-space representations of
an abstract tensor operator $\bfsfL(\omega)$ operating on the
vector field ${\bf E}_{0}(\omega)$ and on the Green function
$\bfsfG_{0}(\omega)$, respectively:
\begin{equation}\label{efieldvacabs}
\bfsfL(\omega)\cdot {\bf E}_{0}(\omega)=0, \qquad
\bfsfL(\omega)\cdot \bfsfG_{0}(\omega)=\openone\otimes\bfsfI.
\end{equation}
The identity operator in real space  is denoted by $\openone$ and
it has the property $\langle {\bf r}|\openone|{\bf
r'}\rangle=\delta^{3}({\bf r}-{\bf r'})$; confusion with the unit
tensor $\bfsfI$ should not arise; the $\otimes$ denotes the tensor
product.

In the presence of an inhomogeneous dispersive linear dielectric,
the wave equation for the electric field is modified into
\begin{equation}\label{efielddiel}
 \bfsfL(\omega)\cdot {\bf E}(\omega)  =
  \bfsfV(\omega)\cdot {\bf E}(\omega),
  \end{equation}
where the  frequency-dependent optical potential $\bfsfV$ is
defined in terms of the dielectric function $\varepsilon({\bf
r},\omega)$ as
\begin{equation}\label{Vdefgeneral}
  \langle {\bf r}|\bfsfV(\omega)|{\bf r'}\rangle  =
  -[\varepsilon({\bf r},\omega)-1](\omega/c)^{2}\bfsfI \delta({\bf
  r}-{\bf r'}).
\end{equation}
The delta-function on the right-hand side defines the potential as
a local quantity (which the T-matrix, to be defined shortly, is
not). In other words, this delta function appears for any
potential.

The electric field ${\bf E}_{0}(\omega)$ is modified into ${\bf
E}(\omega)$, and the two fields are related through the
Lippmann-Schwinger (LS) equation
\begin{subequations}\label{LSvectorabstract}
\begin{eqnarray}
{\bf E}(\omega) & = & {\bf E}_{0}(\omega) +
\bfsfG_{0}(\omega)\cdot \bfsfV(\omega) \cdot {\bf E}(\omega)
\label{LSvectorabstractplain} \\
 &= & {\bf E}_{0}(\omega) + \bfsfG_{0}(\omega)\cdot\bfsfV(\omega)\cdot{\bf E}_{0}(\omega) \nonumber \\
&&+\bfsfG_{0}(\omega)\cdot\bfsfV(\omega) \cdot
\bfsfG_{0}(\omega)\cdot\bfsfV(\omega)\cdot {\bf E}_{0}(\omega)
+\ldots \label{iterativesolu} \\ & = & {\bf E}_{0}(\omega) +
\bfsfG_{0}(\omega)\cdot \bfsfT(\omega) \cdot {\bf
E}_{0}(\omega).\label{LSTvectorabstract}
\end{eqnarray}
\end{subequations}
One can check that indeed the field ${\bf E}(\omega)$ that
satisfies Eq.~(\ref{LSvectorabstractplain}) is also a solution of
Eq.~(\ref{efielddiel}). The solution of
Eq.~(\ref{LSvectorabstractplain}) can be found iteratively in
higher and higher orders of the optical potential ${\bfsfV}$, as
given by the multiple-scattering series in
Eq.~(\ref{iterativesolu}); the (dyadic) T-matrix in
Eq.~(\ref{LSTvectorabstract}) by definition is the formal sum of
the infinite summation in Eq.~(\ref{iterativesolu}). The T-matrix
is a $3\times 3$ tensor. By combining Eqs.~(\ref{iterativesolu})
and (\ref{LSTvectorabstract}), the formal solution for the
T-matrix is
\begin{equation}\label{tsolformal}
\bfsfT(\omega) = \bfsfV(\omega)\cdot\left[\openone\otimes\bfsfI -
\bfsfG_{0}(\omega)\cdot \bfsfV(\omega)\right]^{-1}.
\end{equation}
The scattering problem is solved exactly once the T-matrix is
known.

There may exist optical modes
 that are bound to the scatterer.  Such bound modes correspond to
solutions of  the  LS equation~(\ref{LSvectorabstractplain})  in
the absence of an incident field;  with Eq.~(\ref{tsolformal}) we
can rewrite this homogeneous equation  as
\begin{equation}\label{Tcorrespondstobound}
\bfsfT^{-1}(\omega)\cdot\bfsfV(\omega)\cdot {\bf E}(\omega) = 0.
\end{equation}
It follows that bound solutions of the electric field will
correspond to the poles of the T-matrix. Actually,
Eq.~(\ref{LSTvectorabstract}) also shows that a nonzero solution
for ${\bf E}(\omega)$ can only occur when $\bfsfT(\omega)$ has a
pole. The T-matrix not only solves the scattering problem for
incident fields but also contains all information about bound
modes.

In the presence of the dielectric the Green function also changes,
from $\bfsfG_{0}$ to $\bfsfG$. The latter satisfies
\begin{equation}\label{gformaldiel}
\left[\bfsfL(\omega) - \bfsfV(\omega) \right]\cdot \bfsfG(\omega)
= \openone \otimes \bfsfI,
\end{equation}
 The solution for the Green function  analogous to
 Eq.~(\ref{LSvectorabstract}) for the electric field is the
 three-dimensional Dyson-Schwinger equation
\begin{subequations}\label{DSdefandsolved}
\begin{eqnarray}
\bfsfG(\omega) & = & \bfsfG_{0}(\omega) + \bfsfG_{0}(\omega)\cdot
\bfsfV(\omega) \cdot \bfsfG(\omega) \label{DSvectorabstract} \\
& = & \bfsfG_{0}(\omega)+ \bfsfG_{0}(\omega)\cdot
\bfsfT(\omega)\cdot \bfsfG_{0}(\omega).\label{gintermsoft}
\end{eqnarray}
\end{subequations}
It can be verified that a solution of (\ref{DSvectorabstract})
also is a solution of  Eq.~(\ref{gformaldiel}). The problem how to
find such a solution is solved once the T-matrix
(\ref{tsolformal}) is determined, because an iteration of
Eq.~(\ref{DSvectorabstract}) analogous to the series expansion
(\ref{iterativesolu}) for the electric field shows that the Green
function can also be expressed in terms of the T-matrix, as given
by Eq.~(\ref{gintermsoft}).

Equation (\ref{DSdefandsolved}) also holds when the total
potential $\bfsfV(\omega)$ is a sum of single-scatterer potentials
$\bfsfV_{\alpha}(\omega)$. By iterating one finds that the total
T-matrix for an arbitrary number $N$ of these scatterers is
\begin{equation}\label{tgeneralvecN}
\bfsfT^{(N)} = \sum_{\alpha=1}^{N} \bfsfV_{\alpha} +
\sum_{\alpha,\beta} \bfsfV_{\beta}\cdot \bfsfG_{0}\cdot
\bfsfV_{\alpha} + \sum_{\alpha,\beta,\gamma} \bfsfV_{\gamma}\cdot
\bfsfG_{0}\cdot \bfsfV_{\beta}\cdot \bfsfG_{0}\cdot
\bfsfV_{\alpha} + \dots.
\end{equation}
Often it is  more convenient make an equivalent expansion in terms
of the single-scatterer T-matrices \cite{Gonis00}:
\begin{eqnarray}\label{Tgeneralintsingles}
\bfsfT^{(N)} & = & \sum_{\alpha=1}^{N} \bfsfT_{\alpha} +
\sum_{\alpha,\beta(\ne\alpha)}\bfsfT_{\beta}\cdot \bfsfG_{0}\cdot
\bfsfT_{\alpha} \nonumber \\ & + &
\sum_{\alpha,\beta(\ne\alpha),\gamma(\ne\beta)}\bfsfT_{\gamma}\cdot
\bfsfG_{0} \cdot\bfsfT_{\beta}\cdot \bfsfG_{0}\cdot
\bfsfT_{\alpha} + \dots.
\end{eqnarray}
The frequency dependence was suppressed in
Eqs.~(\ref{tgeneralvecN}) and (\ref{Tgeneralintsingles}). The form
Eq.~(\ref{Tgeneralintsingles}) of the total T-matrix will be used
later in this paper, for model systems where the infinite
summation can be performed explicitly.

\section{Plane scatterers for vector waves}\label{planescatforvec}

The general results of multiple-scattering theory that were
presented in Sec.~\ref{multiplevector} will now be applied to
dielectrics that can be described as a collection of parallel
planes. A suitable representation is chosen, and specific forms of
the potential $\bfsfV$, the free-space Green function
$\bfsfG_{0}$, and the incoming electric field ${\bf E}_{0}$ are
determined. With this,  T-matrices for a single plane and for an
arbitrary number of planes are derived.

\subsection{Dyadic Green function in plane
representation}\label{greenoplane}

A solution for the free-space dyadic Green function can  be found
in three-dimensional Fourier space. By translational invariance,
 $\langle {\bf k}| \bfsfG_{0}(\omega)|{\bf k'}\rangle$
 must be equal to $(2
\pi )^{3} \delta^{3}({\bf k}-{\bf k'}) \bfsfG_{0}({\bf
k},\omega)$. The Green function $\bfsfG_{0}({\bf k},\omega)$
satisfies
\begin{equation}\label{g0four}
\bigl\{ \left[ (\omega/c)^{2}-k^{2} \right] \bfsfI + k^{2}
\hat{\bf k}\hat{\bf k} \bigl\}\cdot \bfsfG_{0}({\bf k},\omega) =
\bfsfI.
\end{equation}
Here, $\hat{\bf k}$ denotes a unit vector in the direction of the
wave vector ${\bf k}$. Equation~(\ref{g0four}) is a $3\times 3$
matrix equation whose representation diagonalizes in the
polarization basis $\{\hat{\bf k},\hat{\bm \sigma}_{1}, \hat{\bm
\sigma}_{2}\}$ with the longitudinal direction $\hat{\bf k}$ and
two orthogonal transverse directions $\hat{\bm \sigma}_{1,2}$. The
solution of (\ref{g0four}) is
\begin{equation}\label{g0componentsfour}
 G_{0}^{jj}({\bf k},\omega)  = \frac{1}{(\omega/c)^{2}-k^{2}},
 \qquad G_{0}^{\hat{\bf
k}\hat{\bf k}}({\bf k},\omega)  =  (c/\omega)^{2},
\end{equation}
where $j$ denotes $\sigma_{1}$ or $\sigma_{2}$. All six
non-diagonal elements of the Green tensor are zero in this
representation. This is the retarded Green function once we assume
that the frequency $\omega$ has an infinitesimally small positive
imaginary part.

The above Fourier representation is not what we need. It is
convenient to work in the ``plane representation'': in
two-dimensional Fourier space in the directions parallel to the
planes and in real space in the $\hat{\bf z}$-direction
perpendicular to the planes. For the polarization representation
choose the orthonormal basis $\{\hat{\bf s}_{{\bf k}},\hat{\bf
v}_{{\bf k}}, \hat{\bf z}\}$. Here, $\hat{\bf z}$ is the unit
vector in the z-direction; $\hat{\bf v}_{{\bf k}}$ is the unit
vector in the direction of the projection of the wave vector ${\bf
k}$ on the plane, so that the wave vector ${\bf k}$ has a
component $k_{\parallel}$ in the $\hat{\bf v}_{{\bf k}}$-direction
and its full representation is $(0,k_{\parallel},k_{z})$; the
$s_{\bf k}$-polarization direction is orthogonal to the optical
plane that is spanned by the other two basis vectors. Then the
operator $\bfsfL(\omega)$ has the form $\langle {\bf
k}_{\parallel}, z|\bfsfL(\omega)|{\bf k'}_{\parallel}, z'\rangle =
(2\pi)^{2}\delta^{2}({\bf k}_{\parallel}-{\bf
k'}_{\parallel})\delta(z-z') \bfsfL({\bf k}_{\parallel},
z,\omega)$, where the operator $\bfsfL({\bf k}_{\parallel},
z,\omega)$ has the matrix representation
\begin{equation}\label{Lmixed}
\left(
\begin{array}{ccc}
(\omega/c)^{2}-k_{\parallel}^{2}+\partial_{z}^{2} & 0 & 0
\\ 0 & (\omega/c)^{2}+\partial_{z}^{2} & -i
k_{\parallel}\partial_{z} \\ 0 & -i k_{\parallel}\partial_{z} &
(\omega/c)^{2}-k_{\parallel}^{2}
\end{array}
\right).
\end{equation}
The Green function in the same representation becomes $\langle
{\bf k}_{\parallel}, z| \bfsfG_{0}(\omega)|{\bf
k'}_{\parallel},z'\rangle = (2\pi)^{2}\delta^{2}({\bf
k}_{\parallel}-{\bf k'}_{\parallel}) \bfsfG_{0}({\bf
k}_{\parallel}, z, z',\omega)$, and the transformed
Eq.~(\ref{g0four}) is a system of differential equations:
\begin{equation}\label{g0eqmixed}
\bfsfL({\bf k}_{\parallel}, z,\omega) \left(
\begin{array}{ccc}
G_{0}^{ss} & G_{0}^{sv} & G_{0}^{sz} \\ G_{0}^{vs} & G_{0}^{vv} &
G_{0}^{vz} \\ G_{0}^{z s} & G_{0}^{zv} & G_{0}^{z z}
\end{array}
\right) = \delta(z-z') \left(
\begin{array}{ccc}
1 & 0 & 0 \\ 0 & 1 & 0 \\ 0 & 0 & 1
\end{array}
\right).
\end{equation}
The $G_{0}^{p q}$ are the components of $\bfsfG_{0}$ and their
arguments $({\bf k}_{\parallel}, z, z',\omega)$ were dropped for
brevity. By choosing the plane representation, the matrix elements
of $\bfsfG_{0}$ only depend on the magnitude and not on the
orientation of ${\bf k}_{\parallel}$. All components involving an
$s$-label are zero, except the $ss$-component.  $G_{0}^{ss}$
satisfies the same differential equation as the Green function
$g_{0}$ of the Helmholtz equation for scalar waves, so that for
$\omega>0$ we have
\begin{equation}\label{g0ss}
G_{0}^{ss}({\bf k}_{\parallel}, z, z', \omega) = g_{0}({\bf
k}_{\parallel}, z, z', \omega)= \frac{ e^{i k_{z}|z-z'|}}{2 i
k_{z}}.
\end{equation}
The variable $k_{z}$ is not independent from ${\bf
k}_{\parallel}$, but rather an abbreviation for
$[(\omega/c)^{2}-k^{2}_{\parallel}]^{1/2}$. The remaining coupled
differential equations of Eq.~(\ref{g0eqmixed}) can also be solved
(again for $\omega>0$), now that $G_{0}^{ss}$ is known:
\begin{subequations}\label{g0comps}
\begin{eqnarray}
G_{0}^{vv}({\bf k}_{\parallel}, z, z', \omega) & = &
\frac{k_{z}^{2}}{(\omega/c)^{2}}g_{0}  \label{g0comps_a}\\
G_{0}^{vz}({\bf k}_{\parallel}, z, z', \omega) & = &
-\frac{k_{\parallel} k_{z}}{(\omega/c)^{2}} g_{0}\;
\mbox{sign}(z-z')\label{g0comps_b}
\\  G_{0}^{zv}({\bf k}_{\parallel}, z, z',
\omega) & = & G_{0}^{vz} \label{g0comps_c} \\ G_{0}^{z z }({\bf
k}_{\parallel}, z, z', \omega) & = &
\frac{1}{(\omega/c)^{2}}\left[k_{\parallel}^{2}g_{0} +
\delta(z-z')\right].\label{g0comps_d}
\end{eqnarray}
\end{subequations}
Green functions in the right-hand sides are understood to have the
arguments $({\bf k}_{\parallel}, z, z', \omega)$.
 The above method of solving differential
equations does not give a value for the sign-function when $z$ is
equal to $z'$. The Green function components (\ref{g0comps}) can
alternatively be found from an inverse Fourier transformation
\begin{equation}\label{fourierbetweengs}
\bfsfG_{0}({\bf k}_{\parallel}, z, z',\omega) = \frac{1}{2 \pi}
\int_{-\infty}^{\infty}\mbox{d}k'_{z}\; \bfsfG_{0}({\bf
k}_{\parallel},k'_{z},\omega)\; e^{ik'_{z}(z-z')}.
\end{equation}
This integration can only be performed in a representation that
does not co-rotate with $k'_{z}$. The basis of
Eq.~(\ref{g0componentsfour}) is not adequate, but again the basis
$\{\hat{\bf s}_{\bf k}, \hat{\bf v}_{\bf k},\hat{\bf z}\}$ suits
well.  With this Fourier method one finds the value 0 for the
sign-function in Eq.~(\ref{g0comps_b}) when $z$ equals $z'$: for
$z=z'$ the relevant integrands in Eq.~(\ref{fourierbetweengs}) are
antisymmetric in the variable $k'_{z}$.

\subsection{Regularization of the Green function}\label{planeregularization}
   The T-matrix of a plane scatterer for vector waves
can be found  by solving the appropriate Lippmann-Schwinger
equation~(\ref{LSvectorabstract}). A plane wave  incident from
$z=-\infty$ with wave vector ${\bf k}$ and arbitrary amplitude
$E_{0}$ and transverse polarization vector $\bfsigma_{\bf
k}=(\sigma_{s},\sigma_{v}, \sigma_{z})$ is scattered by  a plane
at $z=z_{\alpha}$. Because of the symmetry in the in-plane
directions, it is convenient to choose the plane representation
for the LS equation. In terms of the Dirac-notation, the electric
field is a ``ket'';  the plane representation is found by taking
the inner product of Eq.~(\ref{LSvectorabstract}) for the electric
field with the ''bra'' $\langle {\bf k}_{\parallel},z|$, and by
inserting the unit operator
\begin{equation}\label{planerepresentation}
\frac{1}{(2\pi)^{2}}\int\mbox{d}^{2}{\bf
k'}_{\parallel}\mbox{d}z'\; |{\bf k'}_{\parallel},z'\rangle\langle
{\bf k'}_{\parallel},z'|
\end{equation}
at the positions of the dots in the representation-independent
equation~(\ref{LSvectorabstract}). The incident field takes the
form  ${\bf E}_{{\bf k}\bfsigma, {0}}({\bf
k}_{\parallel},z,\omega) = E_{0}\bfsigma_{\bf k} \exp(i k_{z} z)$.
The solution of the LS equation corresponding to this incident
field is  ${\bf E}_{{\bf k}\bfsigma}(\omega)$.    The LS equation
in the mixed representation becomes
\begin{eqnarray}\label{lsvecplane}
{\bf E}_{{\bf k}\bfsigma}({\bf k}_{\parallel}, z,\omega)  =
E_{0}\bfsigma_{\bf k} e^{i k_{z} z}\qquad\qquad\qquad\qquad\qquad && \\
\qquad+ \int_{-\infty}^{+\infty}\mbox{d}z'\;\bfsfG_{0}({\bf
k}_{\parallel}, z,z',\omega)\cdot \bfsfV(z', \omega)\cdot {\bf
E}_{{\bf k}\bfsigma}({\bf k}_{\parallel}, z',\omega).&&\nonumber
\end{eqnarray}
A plane is assumed to be infinitely thin and it can be described
by the optical potential $\bfsfV(z, \omega)=
V(\omega)\delta(z-z_{\alpha})\bfsfI$. (A specific model potential
will be chosen in Sec.~\ref{opticalpotvec}.) The integral can be
evaluated immediately and we get
\begin{eqnarray}\label{LSzisalpha}
{\bf E}_{{\bf k}\bfsigma}({\bf k}_{\parallel},z, \omega) & = &
E_{0}{\bfsigma_{\bf k}} e^{i k_{z} z}  \\ &+ & V(\omega)
\bfsfG_{0}({\bf k}_{\parallel}, z, z_{\alpha}, \omega)\cdot {\bf
E}_{{\bf k} \bfsigma}({\bf k}_{\parallel},z_{\alpha}, \omega).
\nonumber
\end{eqnarray}
The usual way to solve this equation would be to put the position
$z$ equal to $z_{\alpha}$ and to solve for ${\bf E}_{{\bf k}
\bfsigma}({\bf k}_{\parallel},z_{\alpha}, \omega)$. The result
would then be  inserted back  into the above equation to obtain an
expression for ${\bf E}_{{\bf k} \bfsigma}({\bf k}_{\parallel},z,
\omega)$.

However, the Green tensor $\bfsfG_{0}$ is not defined when the
positions $z$ and $z_{\alpha}$ are identical, because of the delta
function in the component  $G_{0}^{zz}$ [Eq.~(\ref{g0comps_d})].
 One could just neglect the delta function,
 as might be correct in other situations \cite{Schwinger78}, but it will
 be argued in Sec.~\ref{regularizeT} that this procedure
would be wrong in our case. Therefore, the Green tensor
(\ref{g0comps}) is not suited for setting up a theory for the
scattering of vector waves by infinitely thin planes.

It is known that ``regularization'' of Green functions is
sometimes needed when modelling finite-sized scatterers as
mathematical objects with zero volume, in order to have a model
that is relevant for optics. (Not always: regularization was not
needed for scalar waves scattering off planes \cite{Wubs02}.) In a
regularization procedure usually a  cutoff parameter is introduced
that modifies the behavior of Green functions at distances much
smaller than optical wavelengths, and mathematical problems are
thus overcome. In some cases, the regularization parameter can be
sent to infinity in the final stage, while in other cases the
cutoff parameter must be kept finite. For example, for point
scatterers the problem of diverging Green functions occurs both
for  scalar and for vector waves. Point scatterers have been
studied extensively and several regularization schemes have been
proposed
 (see \cite{DeVries98a} and references therein).

The same regularization procedure will now be chosen for plane
scatterers as was done before for point scatterers
\cite{DeVries98a}: a high-momentum cutoff is introduced in
three-dimensional Fourier space: instead of the free-space Green
function $\bfsfG_{0}({\bf k}, \omega)$ of
Eq.~(\ref{g0componentsfour}), a regularized free-space Green
function ${\tilde \bfsfG}_{0}({\bf k}, \omega)$ will be used. The
latter is defined in terms of the former as
\begin{equation}\label{g0regfour}
{\tilde \bfsfG}_{0}({\bf k}, \omega) = \left(\frac{\Lambda^{2}}
{\Lambda^{2} + k^{2}}\right) \bfsfG_{0}({\bf k}, \omega).
\end{equation}
The cutoff momentum $\Lambda$ is assumed to be much larger than
the magnitude $\omega/c$ of the optical momentum, so that at
optical wavelengths ${\tilde \bfsfG}_{0}\simeq \bfsfG_{0}$. The
effect of this cutoff in the real-space representation is also
known \cite{DeVries98a}. Here  its effect   on the Green function
in the plane representation is important. After an inverse Fourier
transformation, again only in the $z$-direction, one obtains
(again for $\omega>0$)
\begin{subequations}\label{G0tildecomponents}
\begin{eqnarray}
{\tilde G}_{0}^{ss}({\bf k}_{\parallel}, z, z_{1}, \omega) & = &
\frac{\Lambda^{2}}{\Lambda^{2} + (\omega/c)^{2}} \left( g_{0} +
\frac{e^{-\Lambda_{\parallel}|z-z_{1}|}}{2\Lambda_{\parallel}}
\right)
 \\
{\tilde G}_{0}^{vv}({\bf k}_{\parallel}, z, z_{1}, \omega) & = &
\frac{k_{z}^{2}c^{2}}{\omega^{2}} {\tilde G}_{0}^{ss}
 \\
 {\tilde G}_{0}^{vz}({\bf k}_{\parallel}, z, z_{1},
\omega)& = &  -
\frac{\Lambda^{2}\mbox{sign}(z-z_{1})}{\Lambda^{2}+
(\omega/c)^{2}}\frac{k_{\parallel}}{(\omega/c)^{2}}\times \nonumber\\
& \times & \left[k_{z}g_{0} + (i/2)
e^{-\Lambda_{\parallel}|z-z_{1}|}\right]
  \\
{\tilde G}_{0}^{zv}({\bf k}_{\parallel}, z, z_{1}, \omega)  & = &
{\tilde G}_{0}^{vz}  \\ {\tilde G}_{0}^{zz }({\bf k}_{\parallel},
z, z_{1}, \omega) & = & \frac{\Lambda^{2}}{\Lambda^{2}+
(\omega/c)^{2}} \times  \\ & \times & \left[
\frac{k_{\parallel}^{2}c^{2}}{\omega^{2}}g_{0}  +
\frac{\Lambda_{\parallel}^{2}+(\omega/c)^{2}}
{2\Lambda_{\parallel}(\omega/c)^{2}}
e^{-\Lambda_{\parallel}|z-z_{1}|}\right]. \nonumber
\end{eqnarray}
\end{subequations}
 In the right-hand sides, the arguments
$({\bf k}_{\parallel}, z, z_{1}, \omega)$ of the Green functions
were dropped; $\Lambda_{\parallel}$ is short-hand notation for
$(\Lambda^{2}+k_{\parallel}^{2})^{1/2}$; again, the sign-function
is zero when its argument is.

All components of the regularized Green tensor consist of two
parts: an oscillating and a decaying part, as a function of
$|z-z_{1}|$. The decay occurs at a distance that is a tiny
fraction of an optical wavelength. For $\Lambda |z-z_{1}|\gtrsim
1$, the regularized Green function approaches the unregularized
one. If one would take the limit $\Lambda \rightarrow \infty$,
then all the components in (\ref{G0tildecomponents}) approach the
unregularized components of Eq.~(\ref{g0comps}), and in particular
the limit of the last term in $\bfsfG_{0}^{zz}$ gives the delta
function that made the regularization procedure necessary.
However,  $\Lambda$ is kept finite for the moment, so that
${\tilde G}_{0}^{zz}$  has a finite term that grows with
$\Lambda$. With this result, the Green-function regularization is
complete and a theory of scattering by vector waves from plane
scatterers can be set up.

\subsection{T-matrix of a plane for vector
waves}\label{regularizeT} The regularization entails that  the
Green function is replaced by its regularized version in the LS
equation~(\ref{lsvecplane}). For $z=z_{\alpha}$ that equation
becomes
\begin{eqnarray}\label{LSp}
\left(
\begin{array}{c}
  E_{{\bf k}\bfsigma}^{s}({\bf k}_{\parallel},z_{\alpha},\omega)\\
  E_{{\bf k}\bfsigma}^{v}({\bf k}_{\parallel},z_{\alpha},\omega)\\
  E_{{\bf k}\bfsigma}^{z}({\bf k}_{\parallel},z_{\alpha},\omega)
\end{array}
\right)  =  E_{0} \left(
\begin{array}{c}
\sigma_{s} \\ \sigma_{v} \\ \sigma_{z}
\end{array}
\right) e^{i k_{z}z_{\alpha}} \qquad\qquad\nonumber \\  +
V(\omega) \left(
\begin{array}{ccc}
{\tilde G}_{0}^{ss} & 0 & 0 \\ 0&{\tilde G}_{0}^{vv} &0
\\
0& 0 &{\tilde G}_{0}^{zz}
\end{array}\right)
\left(
\begin{array}{c}
   E_{{\bf k}\bfsigma}^{s}({\bf k}_{\parallel},z_{\alpha},\omega)\\
 E_{{\bf k}\bfsigma}^{v}({\bf k}_{\parallel},z_{\alpha},\omega)\\
  E_{{\bf k}\bfsigma}^{z}({\bf k}_{\parallel},z_{\alpha},\omega)
 \end{array}
 \right).
 \end{eqnarray}
Here ${\tilde G}_{0}^{ss}$ stands for ${\tilde G}_{0}^{ss}({\bf
k}_{\parallel}, z_{\alpha},z_{\alpha},\omega)$, and similarly for
the other components. The off-diagonal elements of the Green
tensor are all zero when the position $z$ is equal to
$z_{\alpha}$, so that the equation  can be solved for every
component separately. By inserting this result into the LS
equation for general $z$, one finds
\begin{eqnarray}\label{solutionfortildeT}
{\bf E}_{{\bf k}\bfsigma}({\bf k}_{\parallel},z, \omega)  =  {\bf
E}_{{\bf k}\bfsigma,{0}}({\bf k}_{\parallel},z,\omega) \nonumber
&& \\  +   {\tilde \bfsfG}_{0}({\bf
k}_{\parallel},z,z_{\alpha},\omega)\cdot {\tilde \bfsfT}({\bf
k}_{\parallel},\omega)\cdot {\bf E}_{{\bf k}\bfsigma,0}({\bf
k}_{\parallel},z_{\alpha},\omega),&&
\end{eqnarray}
where the T-matrix for scattering from a plane by arbitrarily
polarized light is given by
\begin{equation}\label{Tpol}
 {\tilde \bfsfT}({\bf
k}_{\parallel},\omega)= \left(
\begin{array}{ccc}
\frac{V(\omega)}{1 - V(\omega){\tilde G}_{0}^{ss}} & 0 & 0
\\ 0 & \frac{V(\omega)}{1- V(\omega){\tilde G}_{0}^{vv}} & 0 \\
0 & 0 & \frac{V(\omega)}{1- V(\omega){\tilde G}_{0}^{zz}}
\end{array}
\right).
\end{equation}

The scattering of the $s$-polarization component of the light can
be considered independently from the $\hat{\bf v}$ and $\hat{\bf
z}$ directions, according to Eqs.~(\ref{solutionfortildeT}) and
(\ref{Tpol}). It can be verified with
Eqs.~(\ref{G0tildecomponents}, \ref{Tpol}) that since $\Lambda \gg
(\omega/c)$, the matrix component ${\tilde T}^{ss}$ for all
practical purposes is equal to the T-matrix for scalar waves, and
the same holds for the Green tensor component ${\tilde
G}_{0}^{ss}$: the regularization was not necessary for
$s$-polarized light and fortunately it does not affect the
scattering properties of $s$-polarized light.

The need for regularization did show up in the description of
scattering of $p$-polarized light, and there the cutoff might
influence light scattering. Incoming $p$-polarized light is
characterized by its amplitude $E_{0}$, wave vector ${\bf k}$, and
its polarization state $\hat{\bfsigma} = \hat{p} \equiv
(k_{z}/k)\hat{\bf v}_{\bf k}-(k_{\parallel}/k)\hat{\bf z}$.
Written out explicitly, the incoming field is ${\bf E}_{{\bf
k}\bfsigma,{0}}({\bf k}_{\parallel},z,\omega) = E_{0}(0, k_{z}/k,
-k_{\parallel}/k)\;\exp(i k_{z} z)$. For distances far enough from
the plane so that $\Lambda |z-z_{\alpha}|\gg 1$, the term ${\tilde
G}_{0}^{vz}{\tilde T}^{zz}$ in Eq.~(\ref{solutionfortildeT}) falls
off as $\Lambda^{-1}$ and ${\tilde G}_{0}^{zz}{\tilde T}^{zz}$ as
$\exp(-\Lambda_{\parallel}|z-z_{\alpha}|)$, so that for optical
purposes these terms can be neglected. For finite very large
$\Lambda$ we arrive at the following  effective description:
\begin{widetext}
\begin{equation}\label{Eeffectivedescript} \left(
\begin{array}{c}
E^{s}(z) \\E^{v}(z) \\ E^{z}(z)
\end{array}
\right) = \left(
\begin{array}{c}
E_{0}^{v}(z) \\ E_{0}^{v}(z) \\
E_{0}^{z}(z)
\end{array}
\right) + \left(
\begin{array}{ccc}
 G^{ss}_{0} & 0 & 0 \\
0 & G^{vv}_{0} &  G^{vz}_{0}\\ 0& G^{zv}_{0} & G^{zz}_{0}
\end{array}\right)
\left(
\begin{array}{ccc}
T^{ss} & 0 & 0 \\ 0 & T^{vv} & 0 \\ 0 & 0 & 0
\end{array}\right)
\left( \begin{array}{c} E_{0}^{s}(z_{\alpha}) \\
 E_{0}^{v}(z_{\alpha}) \\
E_{0}^{z}(z_{\alpha})
\end{array}
\right),
\end{equation}
\end{widetext}
 where the $ss$-component of the T-matrix is equal to
$V(\omega)\left[1-V(\omega)G_{0}^{ss}\right]^{-1}$, and
analogously for the $vv$-component. The Green functions have
arguments $({\bf k}_{\parallel}, z,z_{\alpha},\omega)$. In this
effective description, - where the T-matrix is denoted by $\bfsfT$
rather than ${\tilde \bfsfT}$ - the cutoff parameter $\Lambda$
does not occur anymore. The cutoff was ne\-ces\-sa\-ry in order to
set up a scattering theory and it shows up in the elements of the
scattering theory such as the T-matrix (\ref{Tpol}) and the
regularized Green function (\ref{G0tildecomponents}). It does not
show up in the electric field, and precisely this enables us to
arrive at the effective description. Note also that  the (large)
value of $G_{0}^{zz}$ has become irrelevant.

The effective description that is obtained here after a
 regularization is different from a theory where the
delta function in $G_{0}^{zz}$ [see Eq.~(\ref{g0comps_d})] would
simply be removed \cite{Schwinger78}. Leaving out the delta
function in the LS equation~(\ref{lsvecplane}) will result in a
nonzero $T^{zz}$, in contrast with Eq.~(\ref{Eeffectivedescript}).
Furthermore, the T-matrix would have the unwanted effect that the
transmitted part of an incoming wave would not be parallel to the
incoming wave. The conclusion is that a regularization of the
Green function was necessary, even when in the end the cutoff
could be sent to infinity.

The equation (\ref{Eeffectivedescript}) defines a true mode of the
electromagnetic field in the presence of a single plane scatterer,
in terms of a  linearly-polarized incoming plane wave with
arbitrary angle of incidence. This is not the complete set of
modes. Other modes, not corresponding to an incoming wave, will be
discussed in Sec.~(\ref{guidmodesvector}), both for a single plane
and for a crystal of planes.

\subsection{Transmission and energy conservation}\label{Opticaltheoremp}
 The transmission of light through the plane can be found
by choosing $z>z_{\alpha}$ in Eq.~(\ref{Eeffectivedescript}). The
transmitted wave can be expressed in terms of the incoming wave as
${\bf E}_{{\bf k}\bfsigma}({\bf k}_{\parallel},z,\omega) =
{\bftau}({\bf k}_{\parallel}, \omega)\cdot {\bf E}_{{\bf
k}\bfsigma, {0}}({\bf k}_{\parallel},z,\omega)$, with the
transmission matrix
\begin{equation}\label{transmissionvector}
{\bftau}({\bf k}_{\parallel}, \omega)= \left(
\begin{array}{ccc}
\tau_{ss}({\bf k}_{\parallel},\omega) & 0 & 0 \\ 0 &
\tau_{vv}({\bf k}_{\parallel},\omega) & 0 \\ 0 & \tau_{zv}({\bf
k}_{\parallel},\omega) & 1
\end{array}
\right),
\end{equation}
which has nonzero elements $\tau_{jj}({\bf k}_{\parallel},\omega)
= \left[ 1 -
V(\omega)G_{0}^{jj}(k_{\parallel},z_{\alpha},z_{\alpha},\omega)\right]^{-1}$
for $j=s,v$. Furthermore, $\tau_{zv}({\bf k}_{\parallel},\omega)=
G^{zv}(k_{\parallel},z_{\alpha},z_{\alpha},\omega)T^{vv}(k_{\parallel},\omega)$.
Both for purely $s$-polarized and for purely $p$-polarized light,
the transmitted electric field is a polarization-dependent scalar
times the incoming electric field vector.

Energy conservation puts a constraint (called `optical theorem')
on the form that the T-matrix of an elastic scatterer can take.
The optical theorem for a plane that scatters scalar waves was
found before \cite{Wubs02}. Since $s$-waves map on scalar waves,
 the optical theorem for the $ss$-component of the T-matrix can be given immediately:
\begin{equation}\label{ssopticaltheorem}
\mbox{Im}\;T^{ss}({\bf k}_{\parallel},\omega) = -\frac{1}{2}
\frac{|T^{ss}({\bf k}_{\parallel},\omega)|^{2}}{k_{z}}.
\end{equation}
The most general T-matrix satisfying this requirement has the form
\begin{equation}\label{mostgeneralformssT}
T^{ss}({\bf k}_{\parallel},\omega)= -\left[F_{s}^{-1}({\bf
k}_{\parallel},\omega)-i/(2k_{z})\right]^{-1},
\end{equation}
where the optical potential $F_{s}({\bf k}_{\parallel},\omega)$
must be a real-valued function.

 For reflection and transmission of $p$-polarized light,
only the matrix element $T^{vv}$ is important. Again, we are
interested in the form that this matrix element can take when
optical energy is conserved. This is the case when the $\hat{\bf
z}$-component of the Poynting vector is the same before and after
the plane. An incoming $p$-polarized plane wave gives the electric
field (\ref{Eeffectivedescript}). With a Maxwell equation the
accompanying magnetic field ${\bf B}$ can also be found. In
SI-units, and in terms of the complex fields ${\bf E}$ and ${\bf
B}$, the cycle-averaged Poynting vector is equal to
$\mbox{Re}[{\bf E}^{*}({\bf r},t)\times {\bf B}({\bf
r},t)]/(2\mu_{0})$ \cite{Loudon83}. When a harmonic wave of
frequency $\omega$ coming from $z=-\infty$ scatters off the plane,
the Poynting vector is
 proportional to $1-(k_{z}c/\omega)^{2}|T^{vv}|^{2}/4$ for
 $(z<z_{\alpha})$. At the other side of the plane one finds
 $|1-ik_{z}(c/\omega)^{2}T^{vv}/2|^{2}$.
By equating the two, the optical theorem for a plane that scatters
$p$-polarized light is found to be
\begin{equation}\label{opticaltheoremppol}
\mbox{Im}\; T^{vv}({\bf k}_{\parallel},\omega)=
\frac{-k_{z}}{2(\omega/c)^{2}}|T^{vv}({\bf
k}_{\parallel},\omega)|^{2}.
\end{equation}
This differs from the optical theorem for $s$-polarized light.
Also, the most general solution of the optical theorem is
different:
\begin{equation}\label{mostgeneralTp}
 T^{vv}({\bf k}_{\parallel},\omega)=-\left[F_{p}^{-1}({\bf
 k}_{\parallel},\omega)- \frac{i k_{z}}{2 (\omega/c)^{2}}\right]^{-1},
 \end{equation}
where the optical potential $F_{p}({\bf
 k}_{\parallel},\omega)$ is real.

\subsection{T-matrix for N planes}\label{TNppol}
Now that the Green function and the T-matrix of a single plane are
known, a multiple-scattering theory can be set up. Assume that
there are $N$  plane scatterers, placed at arbitrary positions.
Assume them to be parallel, so that $s$- and $p$-polarized light
do not mix in the scattering process.

In the general expression~(\ref{Tgeneralintsingles}) for the
T-matrix of a complex dielectric in terms of its simple parts,
Green functions are always sandwiched between T-matrices of
scatterers at different positions. For unequal plane positions
$z_{\alpha}$ and $z_{\beta}$, the value of $\tilde\bfsfG_{0}({\bf
k}_{\parallel},z_{\beta},z_{\alpha},\omega)$ is finite and it can
be taken to be equal to the unregularized $\bfsfG_{0}({\bf
k}_{\parallel},z_{\beta},z_{\alpha},\omega)$, because different
planes are at optical distances apart. Further regularizations are
therefore not required in order to find the $N$-plane T-matrix.

As shown in the Appendix, higher-order terms in the series
(\ref{Tgeneralintsingles}) correspond to higher-order matrix
multiplications of $N\times N$ matrices. The multiplication
property  makes that for parallel planes, the series
(\ref{Tgeneralintsingles}) can be summed exactly. In the Appendix
it is shown how the summation can be done even in the general case
that all planes may have different optical properties, and are
placed at arbitrary non-coinciding positions. Here we specify that
all planes are identical. This gives the central result of this
section, the $N$-plane T-matrix for scattering by vector waves:
\begin{equation}\label{Tpexactlysummed}
\bfsfT^{(N)}(\omega) = \frac{1}{(2\pi)^{2}}\int\mbox{d}^{2} {\bf
k}_{\parallel}\sum_{\alpha,\beta=1}^{N}|{\bf
k}_{\parallel},z_{\alpha}\rangle \bfsfT_{\alpha\beta}^{(N)}({\bf
k}_{\parallel},\omega) \langle {\bf k}_{\parallel},z_{\beta}|
\end{equation}
Each $\alpha\beta$-component $\bfsfT_{\alpha\beta}^{(N)}({\bf
k}_{\parallel},\omega)$ is a $3\times 3$ matrix; the only two
nonzero spatial components  are
\begin{subequations}\label{tNexactcomps}
\begin{eqnarray}
T_{\alpha\beta}^{ss,(N)} & = &
 T^{ss} \left[\bfsfI_{N} -
 \mathcal{W}_{s}T^{ss}\right]_{\alpha\beta}^{-1}  \\
T_{\alpha\beta}^{vv,(N)} & = &
 T^{vv} \left[\bfsfI_{N}
 - \mathcal{W}_{v}T^{vv}\right]_{\alpha\beta}^{-1}
\end{eqnarray}
\end{subequations}
Here, $\bfsfI_{N}$ is the $N\times N$ unit matrix. Arguments
$({\bf k}_{\parallel},\omega)$ were temporarily dropped for
readability. The $N^{2}$ matrix elements
$(\mathcal{W}_{j})_{\alpha\beta}({\bf k}_{\parallel},\omega)$ are
defined as $(1-\delta_{\alpha\beta})G_{0}^{jj}({\bf
k}_{\parallel},z_{\alpha},z_{\beta},\omega)$, for $j=s,v$. The
calculation of $\bfsfT^{(N)}({\bf k}_{\parallel},\omega)$ boils
down to the inversion of an $N\times N$ matrix for the two
transverse polarization directions se\-pa\-ra\-te\-ly.

From now on,  assume that the $N$ planes are placed at regular
distances from each other, with a spacing $a$ between neighbors.
The necessary matrix inversions in Eq.~(\ref{tNexactcomps}) can
then be performed analytically for both polarization directions.
The $s$-wave case maps identically on the situation for scalar
waves, for which the analytical inversion was discussed at length
in \cite{Wubs02}; for $p$-waves the inversion trick goes analogous
and it will not be presented here.

A result from the analytical inversion is that T-matrix elements
and therefore the optical properties of the $N$-plane crystal
strongly depend on the Bloch  wave vectors $K_{s}({\bf
k}_{\parallel},\omega)$ and $K_{p}({\bf k}_{\parallel},\omega)$.
For $p$-polarized light  the Bloch wave vector is given by
$\arccos(C_{p})/a$, with $C_{p} = \cos(k_{z}a) +
C'_{p}\sin(k_{z}a)$; the constant $C'_{p}$ in terms of the
single-plane T-matrix is
\begin{equation}\label{Cpgeneral}
\frac{i k_{z}\left[k_{z}|T^{vv}|^{2}+
2(\omega/c)^{2}\mbox{Im}T^{vv}\right] +
2k_{z}(\omega/c)^{2}\mbox{Re}T^{vv}}{k_{z}^{2}(\mbox{Re}T^{vv})^{2}+
\left[2 (\omega/c)^{2}+ k_{z}\mbox{Im}T^{vv}\right]^{2}}.
\end{equation}
 In general, $C_{p}$
is a complex constant. However, if the optical theorem
(\ref{opticaltheoremppol}) holds, then the imaginary part of
$C_{p}$ becomes identically zero, and  the single-plane T-matrix
will be of the form (\ref{mostgeneralTp}). Likewise, $K_{s}$ is
defined as $\arccos(C_{s})/a$ for a quantity $C_{s}$ that becomes
real when the optical theorem Eq.~(\ref{ssopticaltheorem}) for
$s$-polarized light holds \cite{Wubs02}. In those cases, the
expressions for $C_{s,p}$ become rather simple,
\begin{subequations}\label{Copticaltheorem}
\begin{eqnarray}
C_{s} & = & \cos(k_{z}a) - \left(\frac{F_{s}({\bf
k}_{\parallel},\omega)}{2
k_{z}}\right)\sin(k_{z}a)\label{Csopticaltheorem} \\
C_{p} & = & \cos(k_{z}a) - \left(\frac{k_{z}c^{2}F_{p}({\bf
k}_{\parallel},\omega)}{2
\omega^{2}}\right)\sin(k_{z}a).\label{Cpopticaltheorem}
\end{eqnarray}
\end{subequations}

\subsection{A model for the optical
potential}\label{opticalpotvec}  The most general T-matrices
(\ref{mostgeneralformssT}) and (\ref{mostgeneralTp})    feature as
yet unspecified optical potentials $F_{s,p}$. These should be real
when energy is conserved, but for the rest they can be  arbitrary
functions with  the frequency and the in-plane wave vector as
variables.

 In \cite{Wubs02}, plane scatterers were introduced as
 a simplified model for dielectric slabs of finite thickness $d$ and
 nondispersive dielectric function $\varepsilon(\omega)=\varepsilon$.
The optical potential for the plane scatterer in this model is
obtained via the limiting process of making the thickness $d$ of
the dielectric slab smaller and increasing the polarizability
$(\varepsilon-1)$, while keeping their product constant and equal
to the ``effective thickness'' $D_{\rm eff}$.  (The quantity
$D_{\rm eff}/a$ is called the ``grating strength'' in
\cite{Alvarado01,Zurita02}.) Following the same limiting procedure
as in \cite{Wubs02}, we find the optical potential $F_{s,p}({\bf
k}_{\parallel},\omega) = -V(\omega) = D_{\rm eff}(\omega/c)^{2}$,
identical for the two polarizations. Spatial
 dispersion and anisotropy  would have shown up  in the optical
 potentials
 as a $ k_{\parallel}$- and $\hat{\bf k}_{\parallel}$-dependence,
 respectively. These two phenomena were neglected already as early
as in the wave equation~(\ref{efielddiel}).

In general, $p$-polarized light differs from $s$-polarized light
in that the former will have a Brewster angle at which no light is
reflected from a dielectric interface ($n_{1}\rightarrow n_{2}$).
The Brewster angle $\theta_{B}$ equals  $\tan^{-1}(n_{2}/n_{1})$.
In the limiting procedure for going from a finite slab-in-air to
an infinitely thin plane-in-air, the dielectric contrast
$\sqrt{\varepsilon}/1$ is going to infinity and consequently the
Brewster angle becomes $90^{\circ}$ in that limit. Therefore, in
our limiting procedure, a plane scatterer will not have a Brewster
angle at the same angle as the finite dielectric slab that one
starts out with. In line with this, in \cite{Shepherd97} a
single-plane reflection for $p$-polarized light was determined
that is nonzero for all angles of incidence. The $p$-polarized
propagating modes for a system of plane scatterers will therefore
differ substantially from the corresponding modes in a slab
structure. The absence of a Brewster effect was also noticed in
\cite{Zurita00} where the infinite-crystal version of the
plane-scatterer model is treated.

\section{Optical modes and omnidirectional mirrors}\label{modesvector}
\subsection{Propagating modes}\label{propmodesvector}
The optical modes are the harmonic solutions of the wave
equation~(\ref{efielddiel}). With the solution
(\ref{Tpexactlysummed}) of the  T-matrix, the modes that
correspond to a nonzero incoming plane wave can be given
explicitly as
\begin{eqnarray}\label{modesplanesgen}
{\bf E}_{{\bf k}\bfsigma}({\bf k}_{\parallel}, z,\omega) =
E_{0}\bfsigma_{\bf k} e^{i k_{z} z}\qquad\qquad\qquad\qquad &&
\\+ \sum_{\alpha,\beta} \bfsfG_{0}({\bf k}_{\parallel}, z,
z_{\alpha},\omega)\cdot \bfsfT^{(N)}_{\alpha\beta}({\bf
k}_{\parallel},\omega)\cdot \bfsigma_{\bf k} E_{0} e^{i k_{z}
z_{\beta}}.\nonumber
\end{eqnarray}
These  propagating (or radiative) modes are labelled by the
incoming wave vector ${\bf k}$ and polarization $\bfsigma_{\bf
k}$. The sine of the angle of incidence (with respect to a vector
normal to the planes) is equal to $k_{\parallel}c/\omega$. The
amplitudes of the $s$-polarized modes [with $\bfsigma_{\bf k} =
(1,0,0)$] are identical to the corresponding amplitudes for scalar
waves; the $p$-polarized modes [$\bfsigma_{\bf k} = (0, k_{z}/k,
-k_{\parallel}/k)$] have no scalar analogues.

For  light with wave vector and frequency such that $C_{s,p}>1$
[see Eq.~(\ref{Copticaltheorem})], the Bloch wave vector is
 purely imaginary for the elastic scatterers that we consider.
 Similarly, for $C_{s,p}<-1$,
 the Bloch wave vector equals $\pi$ plus an imaginary number.
 In both situations the light
will feel a stop band, meaning that it will be 100\% reflected
when falling on a semi-infinite system of planes. Otherwise, when
$-1<C_{s,p}<1$, the Bloch wave vector is real and light can
propagate inside the crystal. More will be said about the Bloch
wave vectors later in this section.

\begin{figure*}[t]
\begin{center}
{\includegraphics[width=62mm,height=50mm] {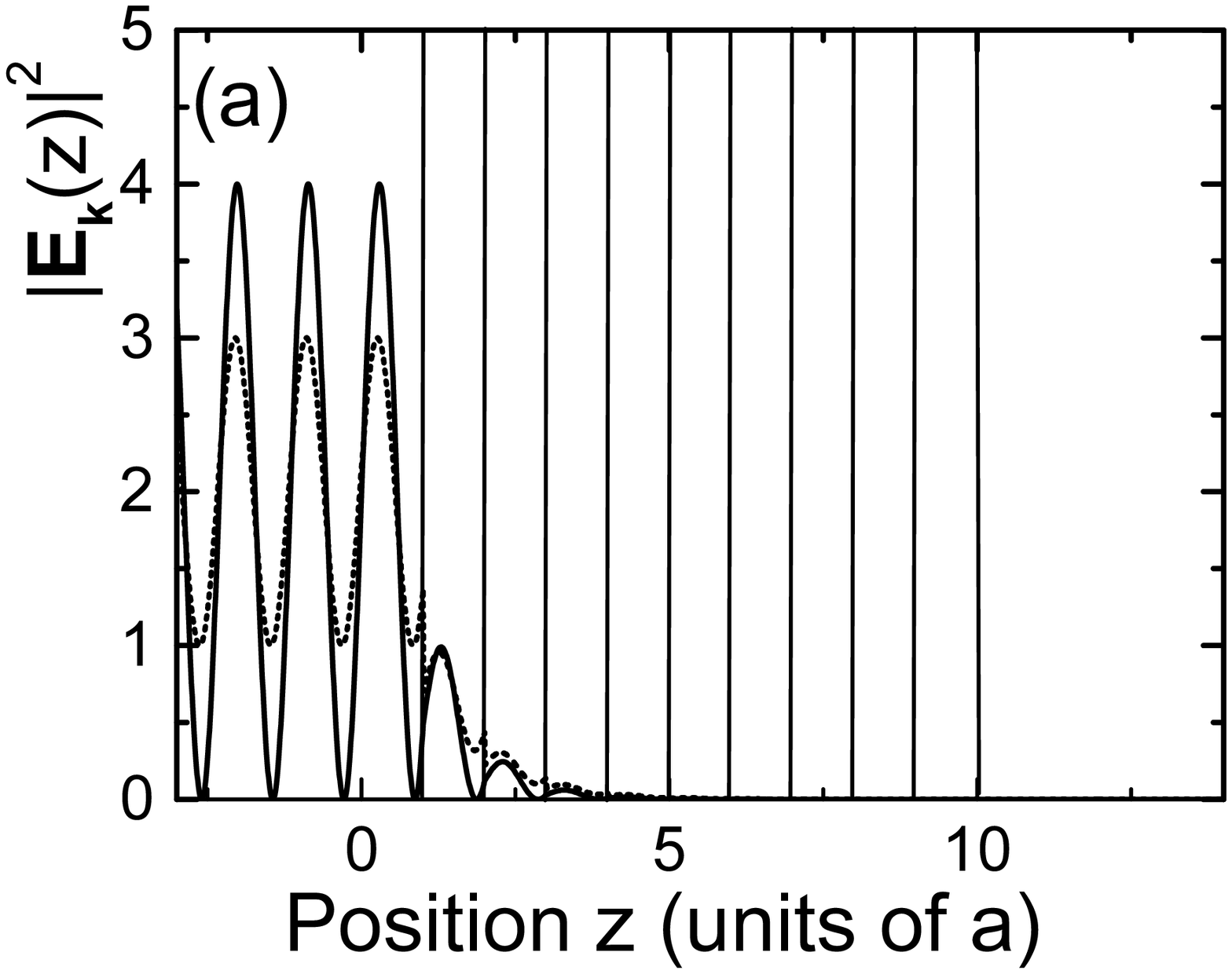}}
{\includegraphics[width=62mm,height=50mm] {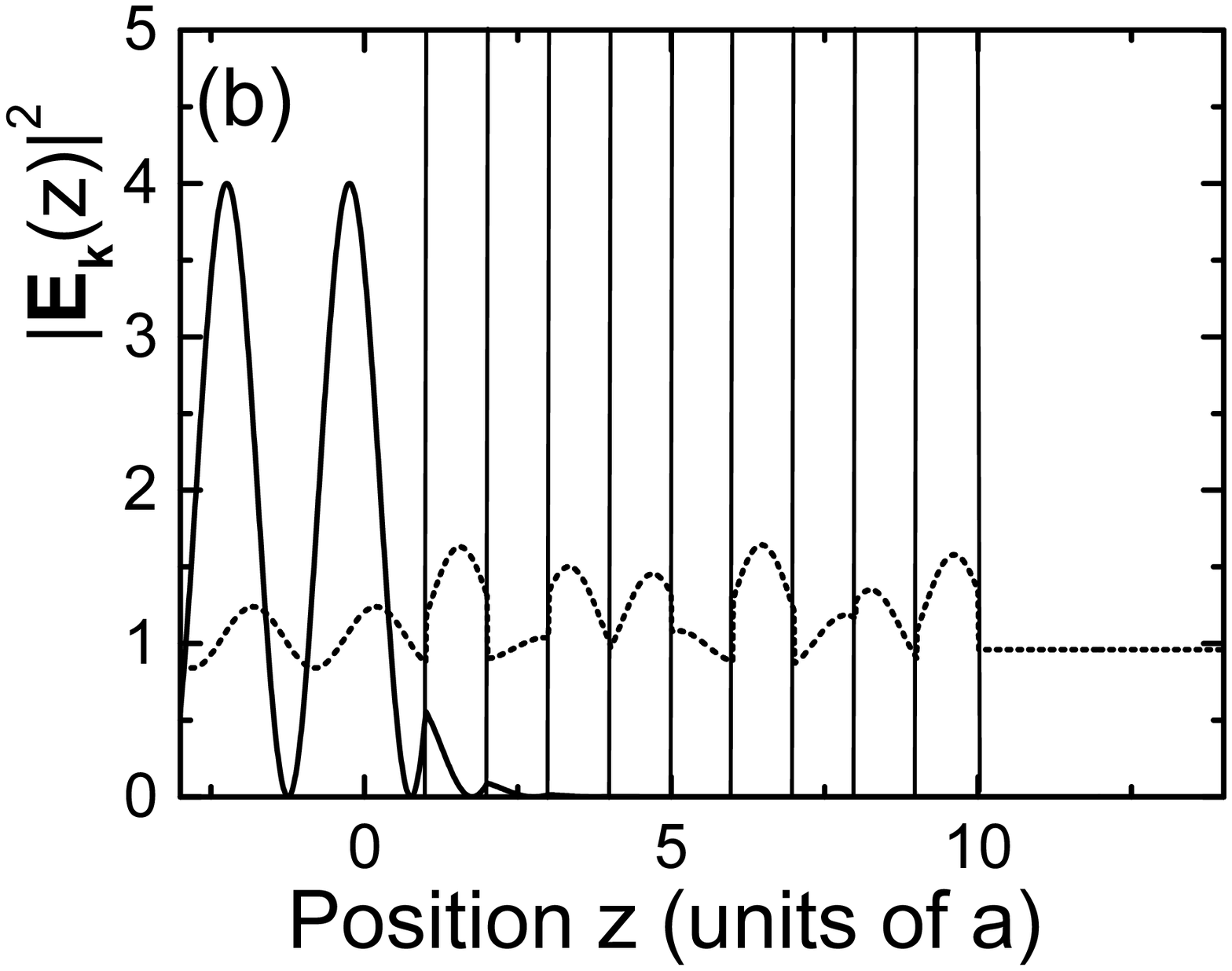}}
\end{center}
\caption{Squares of absolute values of mode functions for
$s$-polarized (solid lines) and $p$-polarized light (dashed
lines), as a function of position. The light is scattered by a
crystal of ten planes with $D_{\rm eff}=0.46a$, separated by a
distance $a$. Both modes correspond to  light incoming from the
left with $a/\lambda=0.5$. Figure (a): $\theta_{\rm
in}=30^{\circ}$; figure (b): $\theta_{\rm
in}=60^{\circ}$.}\label{modessquaredvp}
\end{figure*}

Some plots of mode profiles will now be presented.  Assume that
light comes in from the left. For perpendicularly incident light,
there is no difference between $s$- and $p$-polarization. In
Fig.~\ref{modessquaredvp}
the mode profiles (or squared absolute values of mode functions)
for $s$- and $p$-polarized light inside a ten-plane crystal are
compared both for an incoming angle of $30^{\circ}$ and for
$60^{\circ}$. Fig.~\ref{modessquaredvp}(a) shows that at an angle
of $30^{\circ}$ the mode profiles corresponding to both
polarizations do not differ much yet. Both modes decay rapidly
inside the crystal structure and are reflected (almost)
completely. The Bloch wave vectors are complex for both
polarizations. Only for the $s$-wave the polarization directions
of the incoming and the reflected wave are equal, so that the
amplitude of its mode profile at the left side of the crystal is
four times the amplitude of the incoming electric field.

The situation is different at an incoming angle of $60^{\circ}$,
as shown in Fig.~\ref{modessquaredvp}(b):
 the mode profile of the $s$-polarized light again rapidly
decays inside the crystal (and the corresponding Bloch wave vector
 again has an imaginary part), whereas the $p$-polarized light can propagate
inside the crystal and is transmitted almost completely (and the
Bloch wave vector is real).  For this frequency and incoming
angle, the crystal is a good polarization filter.

 The mode profiles in Figs.~\ref{modessquaredvp}(a,b)
 of the $s$-polarized waves are continuous
whereas $p$-polarized waves  are discontinuous at the positions of
the planes. This reflects the boundary conditions: the tangential
components of the electric fields must be continuous and the
normal components must show a jump at a dielectric interface. The
electric field of $s$-polarized light only has a tangential
component, while $p$-polarized light consists of both tangential
and  normal components. This explains the differences in the  mode
profiles for $s$- and $p$-waves. Notice that in our Green-function
formalism, boundary conditions are automatically satisfied,
whereas in related work based on transfer-matrix methods, boundary
conditions must be considered explicitly
\cite{Visser97,Shepherd97,Zurita02}.

\begin{figure*}[t]
\begin{center}
{\includegraphics[width=60mm,height=50mm]{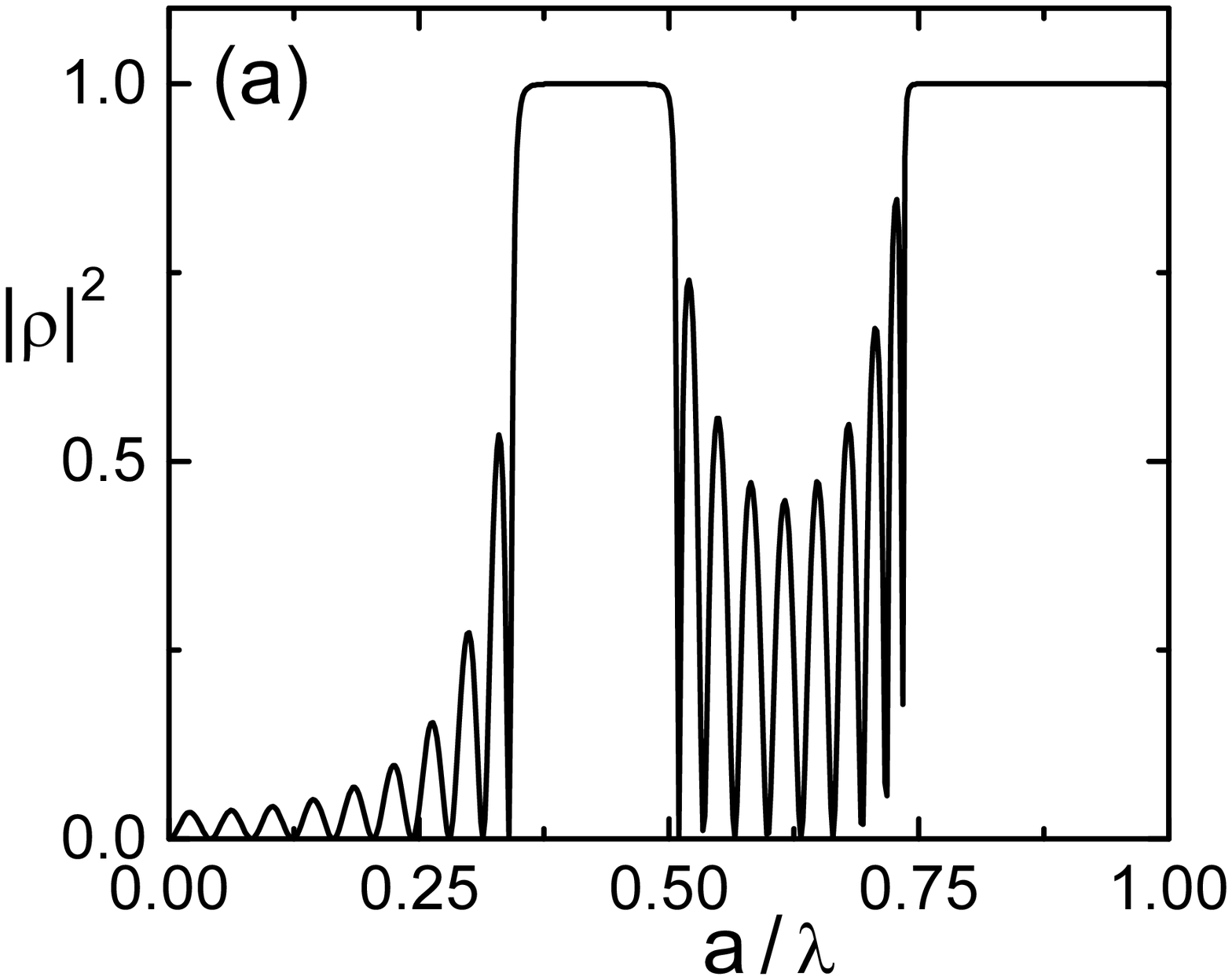}}
{\includegraphics[width=60mm,height=50mm]{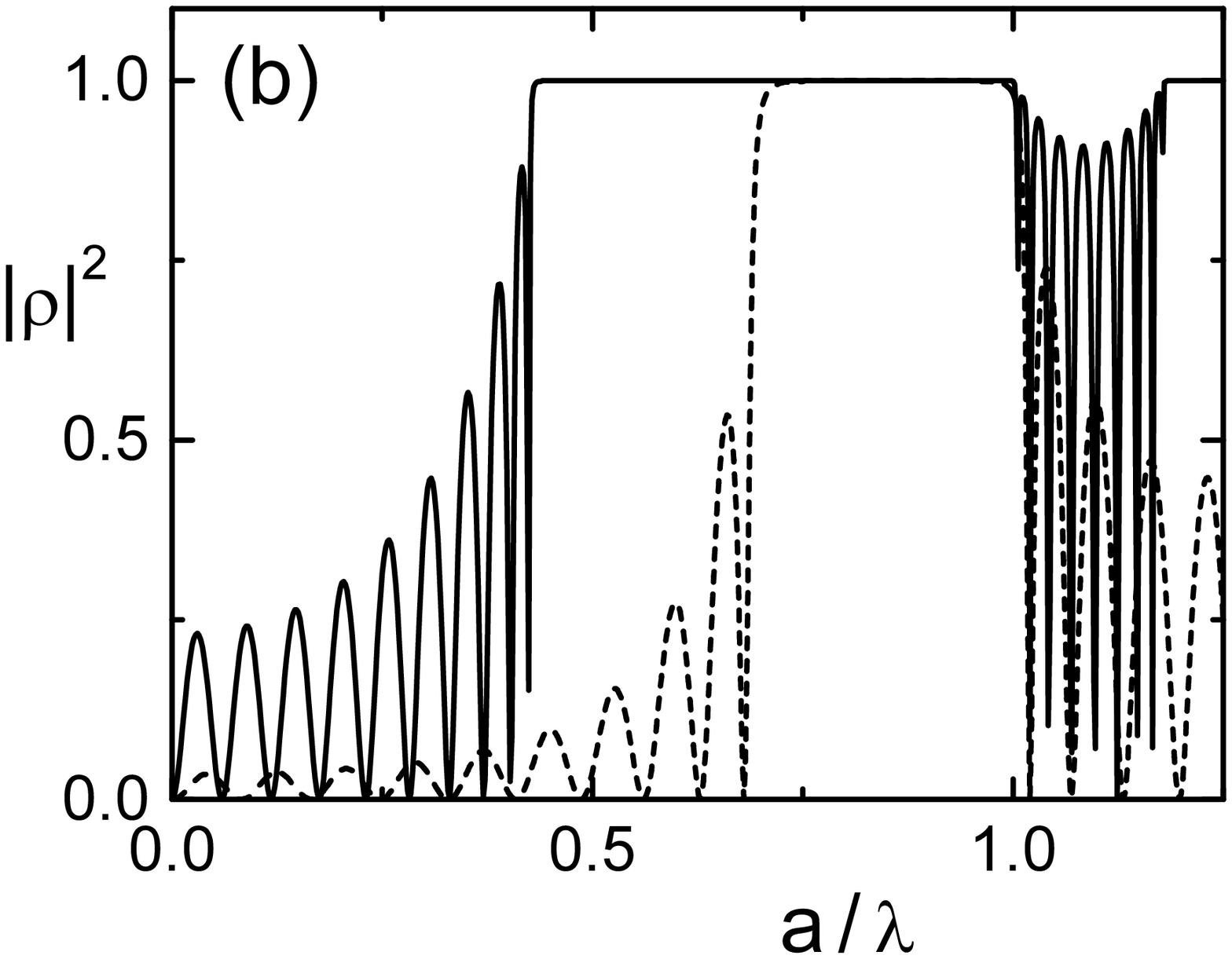}}
\end{center}
 \caption{Reflection off a ten-plane crystal, as
a function of $a/\lambda$, for $s$-polarized light (solid lines)
and $p$-polarized  light (dashed lines). Angles of incidence are
$0^{\circ}$ in (a) and $60^{\circ}$ in (b). The planes have
effective thickness $D_{\rm eff}=0.46a$ and they are separated by
a distance $a$. In (a), the graphs for $s$- and $p$-polarization
overlap.}\label{stopbandsvp}
\end{figure*}

Reflection as a function of frequency by the ten-plane  Bragg
mirror  is plotted in Fig.~\ref{stopbandsvp}
for both polarization directions. The reflection $|\rho|^{2}$
equals $(1-|\tau|^{2})$, with $|\tau|^{2}$ the (relative)
transmitted light intensity. For light incident perpendicularly to
the planes, both transverse polarization vectors are equivalent
and accordingly  in Fig.~\ref{stopbandsvp}(a), the graphs for $s$-
and $p$-polarized light overlap. Differences between the two
polarizations do appear for non-normal incidence. In
Fig.~\ref{stopbandsvp}(b) the  angle of incidence is $60^{\circ}$.
The red edges of the stop bands for $s$-polarized light move to
slightly higher frequencies and the widths of the stop bands
become larger. For $p$-polarized light the red edges of the stop
bands shift to the blue much faster, and the faster so for larger
 angles of incidence.

For scalar waves, a crystal of plane scatterers can be an
omnidirectional mirror \cite{Wubs02}, which means that waves
experience a stop bands for all angles of incidence. For vector
waves, the crystal will only be an omnidirectional mirror if there
are  frequency intervals in which the crystal is an
omnidirectional mirror both for $s$- and $p$-waves.

As stated earlier, it is the Bloch wave vectors $K$ that
distinguish between light that can propagate inside the crystal
(real $K$) and light that feels a stop band (when $K$ is an
imaginary number or $\pi$ plus an imaginary number). In our
formalism, the Bloch wave vectors are the arc cosines of the
constants $C_{s}$ and $C_{p}$ given in
Eq.~(\ref{Copticaltheorem}). As is shown in detail in
\cite{Wubs02}, these Bloch wave vectors show up in expressions for
the $N$-plane T-matrix $T^{(N)}$. It must be said that in the
present T-matrix formalism it is not obvious simply by looking at
the equations that a stop band occurs whenever the Bloch wave
vector has a nonzero imaginary part. Nevertheless, we conclude
from our numerical calculations that the relation does exist. To
give an example, of the four modes in the figures
\ref{modessquaredvp}(a,b), only the $p$-polarized light incoming
at $60^{\circ}$ has a corresponding real Bloch wave vector. It is
an interesting fact that the Bloch wave vector, the role of which
is obvious in infinite crystals, also plays an important role in
finite periodic structures. This was already noticed before in the
context of transfer matrix methods
\cite{Yariv84,Bendickson96,Griffiths00}; here we see the
importance of the Bloch wave vector for finite periodic structures
in a T-matrix formalism.

\begin{figure}[t]
\begin{center}
{\includegraphics[width=65mm]{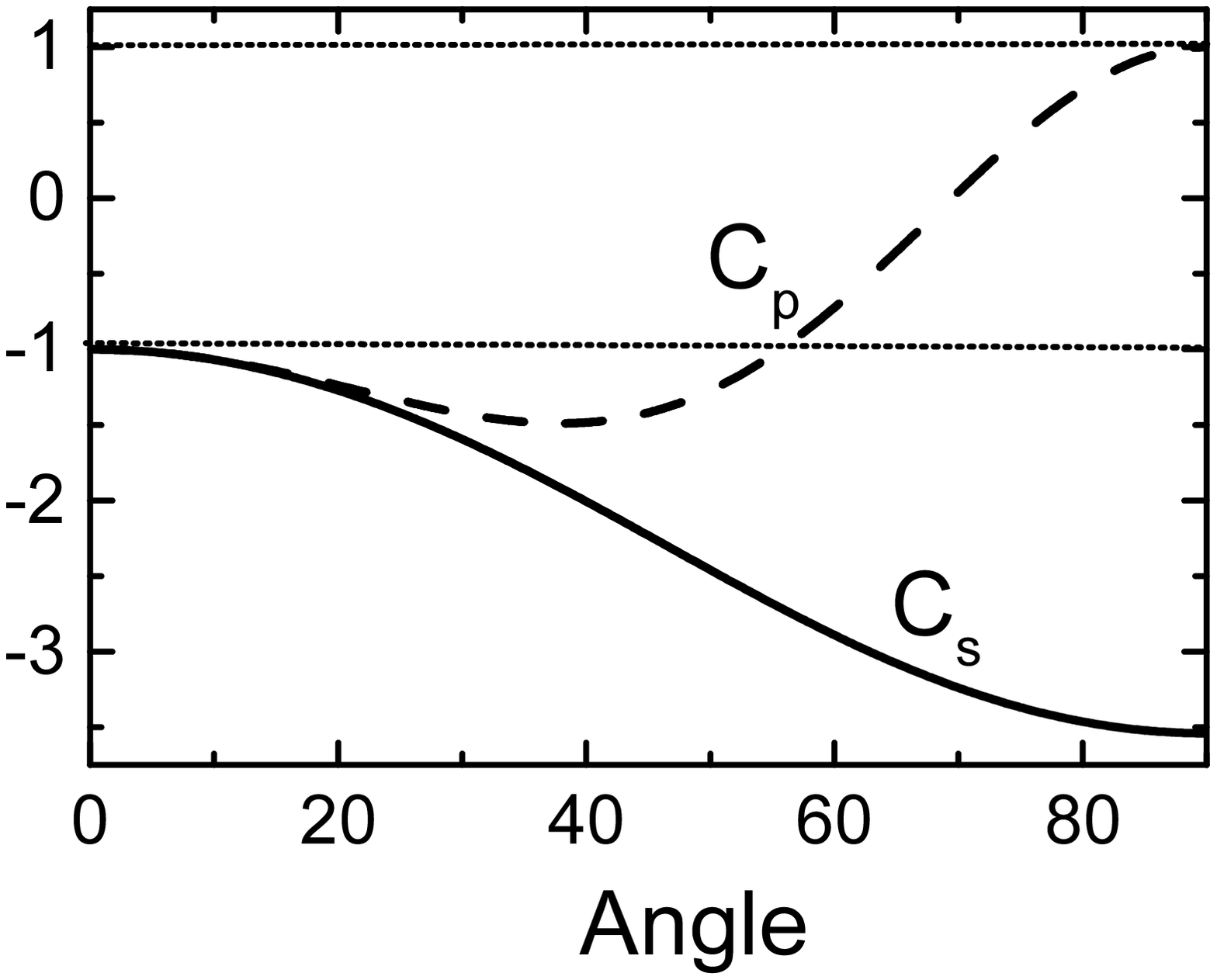}}
\end{center}
\caption{Constants $C_{s}$ and $C_{p}$ as a function of angle of
the incoming light, for the parameters $a/\lambda=0.5$ and $D_{\rm
eff}=0.46 a$. Regions where $-1 \le C \le 1$ correspond to
propagating waves inside the crystal.} \label{cscp046}
\end{figure}

For light of a frequency corresponding to $a/\lambda=0.5$ and
planes with $D_{\rm eff}=0.46$, the  $s$-waves are reflected
omnidirectionally \cite{Wubs02}. In Fig.~\ref{cscp046} we plot
both constants $C_{s}$ and $C_{p}$ for this frequency, as a
function of angle of incidence.
Unlike for $s$-waves, for $p$-waves there are incident angles
larger than the critical angle $\theta_{c} \gtrsim 55^{\circ}$ for
which the values of $C_{p}$ are between -1 and 1. Light incident
with these large angles can propagate inside the crystal and
therefore the crystal is not an omnidirectional mirror for this
frequency. Actually, this information could already be read off
from the mode profile of $p$-polarized light incident at
$60^{\circ}$ in Fig.~\ref{modessquaredvp}(b). The conclusion holds
more generally: for larger $D_{\rm eff}$ the critical angle
$\theta_{c}$ increases, but it can be shown by expanding
Eq.~(\ref{Cpopticaltheorem}) around $\theta_{\rm in}=90^{\circ}$
that for every finite $D_{\rm eff}/a$ and $a/\lambda$ there always
is a finite  interval of angles corresponding to propagating
$p$-polarized light. In conclusion, crystals of identical and
equidistant plane scatterers can reflect vector waves in almost
all (but not in all) directions.

\subsection{Guided modes}\label{guidmodesvector}
Besides propagating modes there can also be bound modes that do
not correspond to incoming light (see Sec.~\ref{multiplevector}).
Bound modes can be found by solving the LS  equation in the
absence of an incoming field. In crystals of plane scatterers,
bound modes are guided modes. They  have imaginary wave vectors in
the $z$-direction and they decay exponentially away from the
planes. Their in-plane wave vectors $k_{\parallel}$ are larger
than $\omega/c$. With each mode, be it of the propagating or
guided type, a nonzero local density of states is associated. In
the following, guided modes will be searched by looking for
nonzero densities of states. (This is a method alternative to
 the one used in \cite{Wubs02} where guided modes
of scalar waves were identified.)

For vector waves, the local optical density of states $N({\bf r},
\omega)$ is defined by \cite{Sprik96}
\begin{equation}\label{LDOSDEF}
-[(2\omega)/(\pi c^{2})]\mbox{Im}\mbox{Tr}\;\bfsfG({\bf r},{\bf
r},\omega),
\end{equation} so it is a scalar
proportional to the trace over the imaginary part of the Green
tensor $\bfsfG({\bf r},{\bf r},\omega)$. In planar geometries the
latter can best be found as an integral over the Green tensor in
the plane representation:
\begin{equation}\label{Gtorealspace}
\bfsfG({\bf r},{\bf r},\omega) =
\frac{1}{(2\pi)^{2}}\int\mbox{d}^{2}{\bf k}_{\parallel}
\bfsfG({\bf k}_{\parallel},z,z,\omega).
\end{equation}
The local density of states can only be nonzero if the imaginary
part of the integrand in (\ref{Gtorealspace}) is nonzero. A guided
mode manifests itself when diagonal elements of this integrand
$\bfsfG({\bf k}_{\parallel},z,z,\omega)$ have  nonzero imaginary
parts for a certain $k_{\parallel}>\omega/c$. For the crystals of
plane scatterers the Green tensor directly follows from the
Dyson-Schwinger equation:
\begin{eqnarray}\label{dsforvector2}
\bfsfG({\bf k}_{\parallel}, z, z',\omega) = \bfsfG_{0}({\bf
k}_{\parallel}, z, z',\omega) \quad\qquad\qquad\qquad\qquad && \\
+ \sum_{\alpha,\beta=1}^{N}\bfsfG_{0}({\bf k}_{\parallel}, z,
z_{\alpha},\omega)\cdot \bfsfT^{(N)}_{\alpha\beta}({\bf
k}_{\parallel},\omega)\cdot \bfsfG_{0}({\bf k}_{\parallel},
z_{\beta}, z',\omega).&& \nonumber
\end{eqnarray}
All three diagonal components of $\bfsfG_{0}({\bf k}_{\parallel},
z, z',\omega)$ become real quantities for
$k_{\parallel}>\omega/c$, and indeed there are no guided modes in
free space. On the other hand, the off-diagonal elements
$G_{0}^{vz}=G_{0}^{zv}$ become purely imaginary when
$k_{\parallel}>\omega/c$ and the latter elements do show up in the
diagonal elements of $\bfsfG$. However, since  they always show up
in paired products, for example in the term
$G_{0}^{zv}T_{\alpha\beta}^{(N),vv}G_{0}^{vz}$, they also give a
real contribution to diagonal elements of $\bfsfG$. The T-matrix
elements are also real when $k_{\parallel}>\omega/c$, except when
the matrix has a pole. Therefore, all guided modes must correspond
to poles of the $s$- or $p$-components of the $N$-plane T-matrix
$\bfsfT^{(N)}$ [see Eq.~(\ref{tNexactcomps})].

First the guided modes of a single plane will be determined. There
can be a guided mode when either $T^{ss}$ or $T^{vv}$ in
Eq.~(\ref{Eeffectivedescript}) has a pole. Now $T^{ss}$ has a pole
when $1-V(\omega)G_{0}^{ss}({\bf
k}_{\parallel},z_{\alpha},z_{\alpha},\omega)$ vanishes. Using the
same model for the optical potential as in
Sec.~\ref{opticalpotvec}, we find the dispersion relation
$\kappa_{1}^{(1)} = D_{\rm eff}(\omega/c)^{2}/2$ for one and only
one guided mode corresponding to $s$-polarized light. Here,
$\kappa$ is the positive square root
$[k_{\parallel}^{2}-(\omega/c)^{2}]^{1/2}$ for $k_\parallel >
\omega/c$. A single-plane guided mode with this dispersion
relation was also found in \cite{Shepherd97}, and in \cite{Wubs02}
for scalar waves.

A pole of $T^{vv}$ occurs when $1-V(\omega)G_{0}^{vv}({\bf
k}_{\parallel},z_{\alpha},z_{\alpha},\omega)$ vanishes, which is
equivalent to the requirement that $D_{\rm eff}\kappa/2$ equals
$-1$. Now in principle, $D_{\rm eff}$ could be negative when
modelling a slab of negative dielectric function as a plane
scatterer. However, in the physical situations that we are
interested in, the effective thickness is real and positive (see
Sec.~\ref{opticalpotvec}). Therefore, there is no guided mode
corresponding to $p$-polarized light for a single plane scatterer.
This result was also derived in \cite{Shepherd97}, where only the
cases of a single plane and infinitely many planes were
considered.

At this point it is worthwhile to compare the guided modes of a
dielectric slab (thickness $d$, dielectric constant $\varepsilon$)
in air with the guided modes of an infinitely thin plane with
effective thickness $D_{\rm eff}= (\varepsilon-1)d$. For the slab,
   the number $M$ of
guided modes is the same for both polarizations in the special
case considered here, and equal to \cite{Urbach98}
\begin{equation}\label{numberofmodes}
M_{s,p} = 1+ \left[2d\sqrt{\varepsilon-1}/\lambda\right].
\end{equation}
Here, $\left[X\right]$ stands for the largest integer smaller than
or equal to $X$. In the large-wavelength limit there is a single
guided mode for each polarization direction. The second guided
mode appears when $a/\lambda=a/(2\sqrt{d D_{\rm eff}})$.  For
example, for $d=0.1a$ and $D_{\rm eff}=0.46a$, a second guided
mode exists when $a/\lambda> 2.3$. We are interested in
frequencies around $a/\lambda=0.5$, where the first stopband for
normally incident light occurs (see Fig.~\ref{stopbandsvp}). For
these frequencies, both the plane scatterer and the dielectric
slab have a single $s$-polarized guided mode; the slab has a
single $p$-polarized guided mode, whereas the plane scatterer has
no such guided mode at all.

Now we determine the guided modes of a finite crystal of $N$
parallel and equidistant planes, using the same Green-function
method as for the single plane. First look for the poles of the
component $T_{\alpha\beta}^{ss,(N)}({\bf k}_{\parallel},\omega)$.
This is easy, because this component is identical to the $N$-plane
T-matrix $T^{(N)}$ for scalar waves, for which it was found that
there are at most $N$ guided modes in a crystal of $N$ planes
\cite{Wubs02}. For an infinite number of planes, the guided modes
form a band, as was found in \cite{Shepherd97,Alvarado01}. Now
look for guided modes corresponding to $p$-polarized light. The
poles of the component $T_{\alpha\beta}^{vv,(N)}({\bf
k}_{\parallel},\omega)$ occur when the determinant $\det [({\bf
T}^{vv,(N)})^{-1}]$ is equal to zero. An expression for this
determinant can be found just like was done for scalar waves in
\cite{Wubs02}. The result is that for $p$-polarized light there
are  guided modes if the following equation has  nontrivial
solutions $\omega(\kappa)$:
\begin{equation}\label{pguidedifzero}
\sin[(N+1)K_{p}a]+ \left[\frac{\kappa F -
2(\omega/c)^{2}}{2(\omega/c)^{2}}\right] e^{-\kappa a}\sin(N
K_{p}a) = 0.
\end{equation}
 The Bloch wave vector $K_{p}$ is still defined as
$a^{-1}$ times the arccosine of $C_{p}$, which reads $C_{p} =
\cosh(\kappa a) + [(\kappa c^{2} F)/(2 \omega^{2})]\sinh(\kappa
a)$ in terms of $\kappa$.

Eq.~(\ref{pguidedifzero}) should lead to the dispersion relations
$\omega(\kappa)$ for the guided modes, if they exist. When
increasing the frequency, new guided modes would appear that at
first are only just captured by the structure so that
$\kappa=0^{+}$. It is therefore convenient to count the guided
modes in the small-$\kappa$ limit. Let the constant $\chi$ be
defined as $a\sqrt{1+Fc^{2}/(a\omega^{2})}$. Then $C_{p}$ can be
written up to second order in $\kappa$ as
\begin{equation}\label{cpestimate}
C_{p}= 1 + (\chi \kappa)^{2}/2 + o(\kappa^{3}) = \cosh(\chi
\kappa) + o(\kappa^{3}).
\end{equation}
 To the
same order in $\kappa$, the Bloch wave vector $K_{p}$ becomes
equal to $i \chi\kappa/a$. Therefore, solutions of
Eq.~(\ref{pguidedifzero}) will only exist when
 $\sinh[(N+1)\chi \kappa]$ equals $\sinh(N \chi \kappa)$,
 or equivalently when $\chi \equiv 0$. Since  $F$ is taken to be
 $D_{\rm eff} (\omega/c)^{2}$,  there are guided modes for $p$-polarized light
whenever $1 + D_{\rm eff}/a=0$. So in  finite crystals of $N$
planes with a positive effective thickness, there are no guided
modes corresponding to $p$-polarized light. This is in agreement
with the result just obtained for the single plane, and with the
results in \cite{Shepherd97,Zurita02} for infinite crystals.

In conclusion, there are at most $N$ guided modes in the finite
crystal of plane scatterers, all modes corresponding to
$s$-polarized light. The comparison of the single plane and the
single slab indicates that $s$-waves in plane scatterers are a
good model for $s$-waves in slabs, at least  for frequencies
around the first stopband. For the $p$-polarized guided modes, the
conclusion must be that in the finite slab structures there are
guided modes which have no analogues in the crystal of plane
scatterers.

\section{Spontaneous emission}\label{SEvector}
\subsection{Application to layered dielectrics}\label{segenerallayers}
In free space, the spontaneous-emission rate $\Gamma_{0}$ of an
atom with dipole moment ${\bm \mu}$ and transition frequency
$\Omega$ equals $\mu^{2}\Omega^{3}/(3\pi\hbar\varepsilon_{0}
c^{3})$. When  embedded in an inhomogeneous dielectric, the rate
$\Gamma$ will in general be different, as can be found with
Fermi's golden rule \cite{Glauber91}
\begin{equation}\label{SEgeneralinhom}
\Gamma({\bm \mu},{\bf R}, \Omega) = \pi
\sum_{l}\frac{\omega_{l}}{\hbar \varepsilon_{0}}|\bfmu\cdot {\bf
E}_{l}({\bf R})|^{2}\;\delta(\omega_{l}-\Omega).
\end{equation}
The ${\bf E}_{l}$ are the normal-mode solutions with
eigenfrequencies $\omega_{l}$ of the wave
equation~(\ref{efielddiel}). The spontaneous-emission rate can
alternatively be expressed in terms of the Green function of the
medium
\begin{equation}\label{SEintermsofG}
\Gamma(\bfmu,{\bf R}, \Omega) = \frac{-2\Omega^{2}}{\hbar
\varepsilon_{0} c^{2}} \mbox{Im}\left[\bfmu\cdot \bfsfG({\bf
R},{\bf R},\Omega)\cdot \bfmu\right].
\end{equation}
(See \cite{AgarwalWylie} for early derivations of this relation;
in \cite{Knoell01}
 a modern derivation is given for inhomogeneous and absorbing dielectric media.)
 In Eq.~(\ref{SEintermsofG}), $\bfsfG$ is the
classical dyadic Green function of the electric-field wave
equation~(\ref{efielddiel}). For homogeneous dielectrics, it is
known that the total Green function is the sum of a transverse
part that describes radiative decay and a longitudinal part
describing nonradiative decay \cite{Barnett96}. Here, the
Eqs.~(\ref{SEgeneralinhom}) and (\ref{SEintermsofG}) are
equivalent, because nonradiative decay is absent for dielectrics
with real dielectric functions.

 Layered
dielectrics (not necessarily plane scatterers) are translation
invariant in two directions, which can be chosen to be the
$(\hat{\bm x},\hat{\bm y})$ directions. Spontaneous-emission rates
will only depend on the $z$-coordinate of the atomic position
${\bf R}=(x,y,z)$. It is then easiest to first calculate the Green
function in the plane representation $\bfsfG({\bf k}_{\parallel},
z, z,\Omega)$.   This Green function must be Fourier transformed
back to real space as in Eq.~(\ref{Gtorealspace}) in order to find
the local Green function of Eq.~(\ref{SEintermsofG}) that
determines spontaneous-emission rates.

A slight complication in doing the
integration~(\ref{Gtorealspace}) is that the plane representation
for $\bfsfG({\bf k}_{\parallel}, z, z,\Omega)$
 is co-rotating with the incoming wave
vector ${\bf k}_{\parallel}$, a variable that must now be
integrated over. A fixed basis $\{\hat{\bf x},\hat{\bf y},\hat{\bf
z}\}$ is needed instead and it is chosen  such that the atomic
dipole becomes $(\mu_{x},0,\mu_{z})$ in the new representation.
Write the two-dimensional integral $\int\mbox{d}^{2}{\bf
k}_{\parallel}$  in polar coordinates as
$\int_{0}^{\infty}\mbox{d}{
k}_{\parallel}k_{\parallel}\int_{0}^{2\pi}\mbox{d}\hat{\bf
k}_{\parallel}$. After doing the angular integral, only diagonal
elements of the dyadic Green function survive.

The total spontaneous-emission rate is the sum of two
contributions, the perpendicular and the parallel decay rate
\begin{subequations}\label{seperp}
\begin{eqnarray}
\Gamma_{z}(z,\Omega) & = &
-\frac{3c\Gamma_{0}\mu_{z}^{2}}{\Omega\mu^{2}}\mbox{Im}
\int_{0}^{\infty}\mbox{d}k_{\parallel} k_{\parallel}
G^{zz} \label{gammaz} \\
\Gamma_{x}(z,\Omega) & = &
-\frac{3c\Gamma_{0}\mu_{x}^{2}}{\Omega\mu^{2}}\mbox{Im}
\int_{0}^{\infty}\mbox{d}k_{\parallel} k_{\parallel} \frac{G^{ss}
+G^{vv}}{2}  \label{gammax}
\end{eqnarray}
\end{subequations}
(Green-function arguments $(k_{\parallel}, z,z,\Omega)$ were again
dropped.) The parallel decay rate has a contribution both from
$s$- and $p$-polarized light (through $G^{ss}$ and $G^{vv}$,
respectively) whereas the perpendicular decay rate only has a
$p$-polarized decay channel (through $G^{zz}$). Notice that the
(real) delta-function term in $G_{0}^{zz}$ does not play a role in
the emission rates. The spontaneous-emission rates in
Eq.~(\ref{gammaz}) and (\ref{gammax}) are integrals over all
possible lengths of the in-plane wave vector. Both rates can be
subdivided into a propagating-mode (or radiative-mode) rate
corresponding to the integration of $k_{\parallel}$ from $0$ to
$\Omega/c$, and a guided-mode rate which is the integral from
$\Omega/c$ to infinity.

\subsection{Spontaneous emission near plane scatterers}
\label{seplanes} The general expressions obtained in
Sec.~\ref{segenerallayers} for spontaneous emission in layered
structures will now be applied to crystals of plane scatterers.
 Combine the general expressions (\ref{seperp})
for spontaneous-emission rates in layered dielectrics with the
Green functions in the plane representation that were determined
in Eq.~(\ref{dsforvector2}) for a crystal of plane scatterers.
Because of the absence of $p$-polarized guided modes, the parallel
decay rate near plane scatterers can be subdivided into three
(instead of four) parts: an $s$-polarized radiative-mode rate
$(s{\rm r})$, a $p$-polarized radiative-mode rate $(p{\rm r})$,
and an $s$-polarized guided-mode rate $(s{\rm g})$. Again, because
of the absence of $p$-polarized guided modes, the perpendicular
decay rate $\Gamma_{z}$ is purely radiative. Here is a list of the
nonzero partial decay rates:
\begin{subequations}\label{allgammas}
\begin{eqnarray}
 \Gamma_{x}^{s{\rm r}}(z,\Omega) & = &
-\frac{3c\Gamma_{0}}{2\Omega}\left(\frac{\mu_{x}}{\mu}\right)^{2}
 \mbox{Im} \int_{0}^{\Omega/c}\mbox{d}k_{\parallel}
k_{\parallel} G^{ss}\label{gammax_sr}
\\ \Gamma_{x}^{p{\rm r}}(z,\Omega) & = &
-\frac{3c\Gamma_{0}}{2\Omega}\left(\frac{\mu_{x}}{\mu}\right)^{2}
\mbox{Im} \int_{0}^{\Omega/c}\mbox{d}k_{\parallel} k_{\parallel}
G^{vv}\label{gammax_pr}
\\ \Gamma_{x}^{s{\rm g}}(z,\Omega) & = &
-\frac{3c\Gamma_{0}}{2\Omega}\left(\frac{\mu_{x}}{\mu}\right)^{2}
\mbox{Im} \int_{\Omega/c}^{\infty}\mbox{d}k_{\parallel}
k_{\parallel} G^{ss}\label{gammax_sg}
\\
\Gamma_{z}(z,\Omega) & = &
-\frac{3c\Gamma_{0}}{\Omega}\left(\frac{\mu_{z}}{\mu}\right)^{2}
\mbox{Im} \int_{0}^{\Omega/c}\mbox{d}k_{\parallel} k_{\parallel}
G^{zz}. \label{gammazplanes}
\end{eqnarray}
\end{subequations}
To be precise, in Eq.~(\ref{gammazplanes}) it was used that the
tensor element $G^{zz}$ has a vanishing imaginary part (leading to
a vanishing contribution to the density of states) for
$k_{\parallel}>\Omega/c$; for the same reason, there is no
guided-mode rate analogous to Eq.~(\ref{gammax_sg}) corresponding
to $G^{vv}$. These properties were found in
Sec.~\ref{guidmodesvector}.

\begin{figure*}[t]
\begin{center}
{\includegraphics[width=62mm,height=50mm]{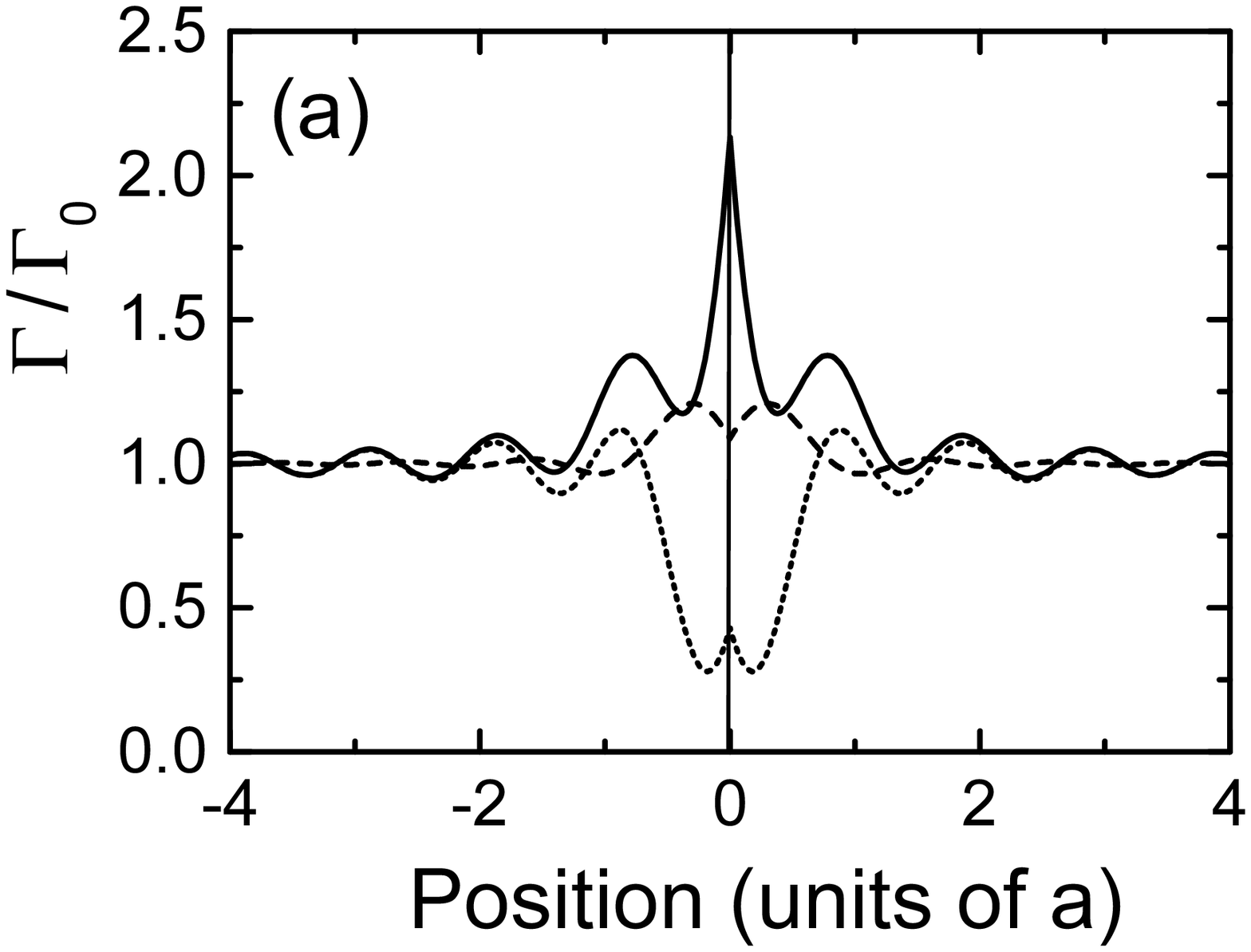}}
{\includegraphics[width=62mm,height=50mm]{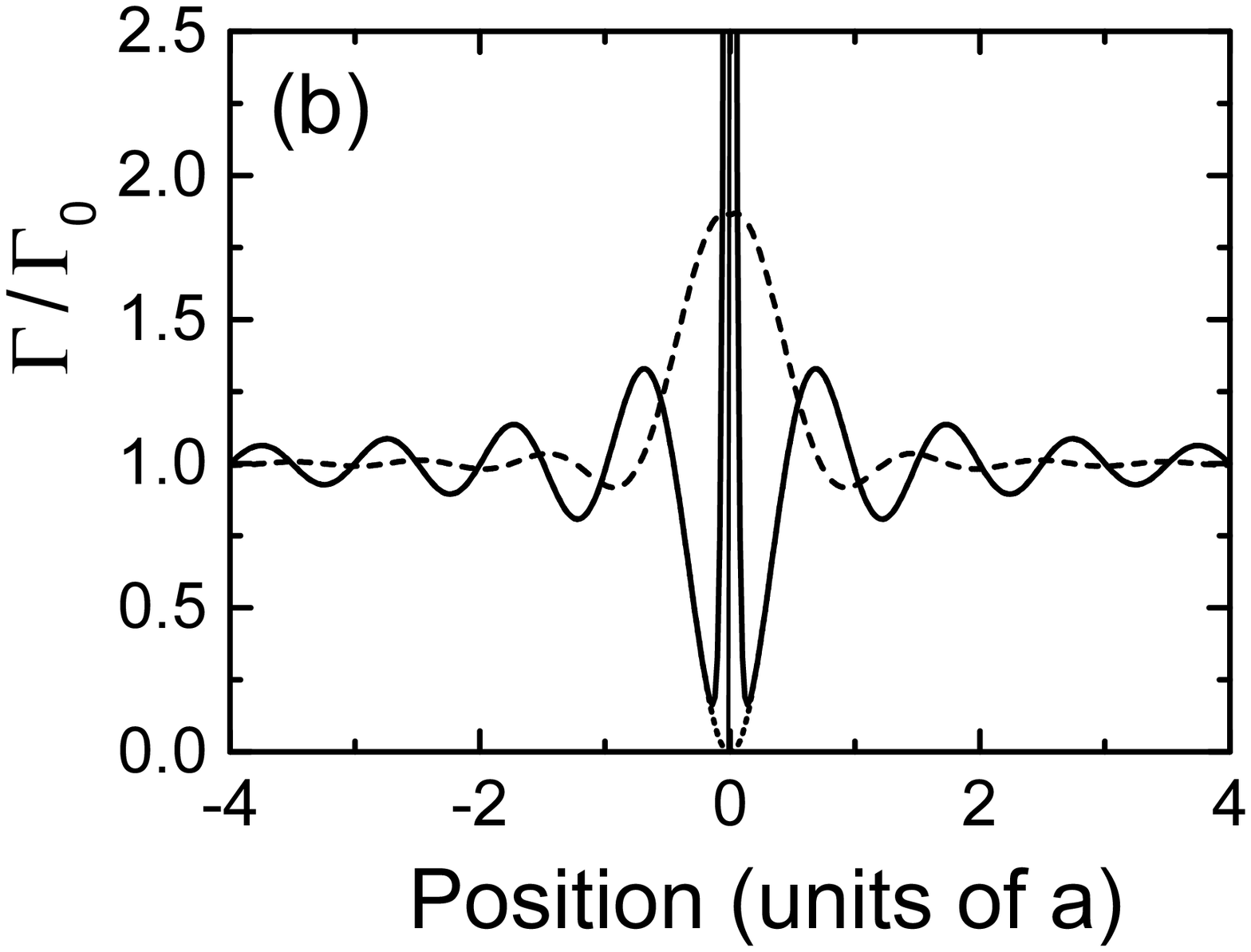}}
\end{center}
\caption{Spontaneous-emission rates of dipoles near  a single
partially transmitting plane scatterer, relative to $\Gamma_{0}$.
The wavelength is chosen such that $a/\lambda=0.5$. Solid lines
correspond to total spontaneous-emission rates $\Gamma_{x}$ for
dipoles parallel to the plane, dotted lines are radiative
contributions to $\Gamma_{x}$, and the dashed lines denote
$\Gamma_{z}$. In (a), the effective thickness $D_{\rm eff}$ of the
plane  equals $0.46 a$ and in (b) $D_{\rm
eff}=10a$.}\label{serates05N1}
\end{figure*}

With all partial emission rates spelled out now, we  first study
spontaneous-emission rates near a single plane, for which the
Green function in the plane representation (\ref{dsforvector2})
features the single-plane T-matrix  of
Eq.~(\ref{Eeffectivedescript}). In Fig.~\ref{serates05N1}(a),
spontaneous-emission rates as a function of position are plotted
for $D_{\rm eff}=0.46 a$.
 For
both orientations of the dipole, far away from the plane the rate
approaches the free-space value. Close to the plane, $\Gamma_{x}$
is larger than $\Gamma_{0}$, but it consists of a rate into
propagating modes that is less than $\Gamma_{0}$, and a
guided-mode rate. Close to the plane, $\Gamma_{z}$ is larger than
$\Gamma_{0}$, but the maximal values of  (the purely radiative)
$\Gamma_{z}$ occur somewhat away from the plane.

The contribution of radiative and guided $s$-waves for an atom
with $\mu = \mu_{x}$ is the same as for scalar waves with ``scalar
dipole moment'' $\mu$, but since the total decay rate $\Gamma_{0}$
for vector waves is larger than for scalar waves, the relative
contributions of $s$-waves to $\Gamma/\Gamma_{0}$ are smaller for
vector waves (by a factor $3/4$).

In Fig.~\ref{serates05N1}(b), the same rates are plotted, this
time for a plane with $D_{\rm eff}=10 a$ that reflects light
almost ideally: near the plane, $\Gamma_{z}$ is almost twice
$\Gamma_{0}$. The maximum values of $\Gamma_{z}$ still occur away
from the plane, although this has become invisible in
Fig.~\ref{serates05N1}(b). The propagating-mode part of
$\Gamma_{x}$ has decreased and is practically zero on the plane.
The partial emission rate into the guided mode has a much larger
(but finite, not shown) amplitude near the plane. The other
prominent difference in the two figures is that the `spike' in the
emission rates due to the guided modes has become much narrower.
Indeed, from Eqs.~(\ref{LDOSDEF}-\ref{dsforvector2}), it follows
that the guided-mode rate decays exponentially away from the plane
like $\exp(-2\kappa_{1}^{(1)}|z|)$. It follows from the dispersion
relation for $\kappa_{1}^{(1)}$ that was obtained in
Sec.~\ref{guidmodesvector},  that an increase in $a/\lambda$ or in
$D_{\rm eff}$ will give narrower spikes.

\begin{figure}[t]
\begin{center}
{\includegraphics[width=65mm, height=60mm]{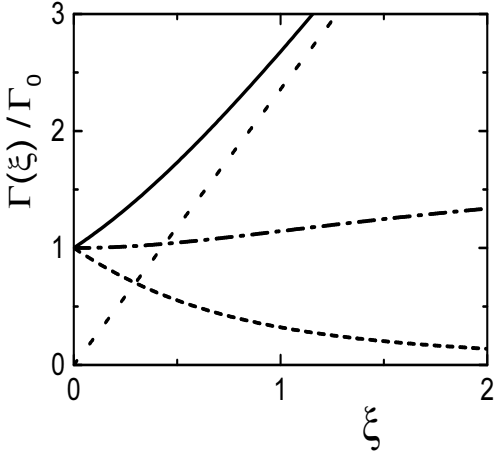}}
\end{center}
\caption{Spontaneous-emission rates of dipoles at the position of
the single plane scatterer, as a function of the dimensionless
parameter $\xi = \pi D_{\rm eff}/\lambda$. The rate $\Gamma_{x}$
(solid line) is the sum of a rate into radiative (dashed line) and
into guided modes (dotted line). The purely radiative-mode rate
$\Gamma_{z}$ is the dash-dotted line.}\label{seratesxi}
\end{figure}

 In the limit that the atomic position $z$ becomes
equal to the plane position $z_{\alpha}=0$, the
spontaneous-emission rates  can be calculated analytically for
both dipole orientations. For a dipole perpendicular to the
planes,
\begin{equation}\label{sezcloseoneplane}
\Gamma_{z}(z_{\alpha},\Omega) = \Gamma_{0}\left(2 + \frac{3}{2
\xi^{2}}\right) - \frac{3}{2}\Gamma_{0}\left(1 +
\frac{1}{\xi^{2}}\right)\frac{\arctan(\xi)}{\xi},
\end{equation}
where the dimensionless parameter $\xi$ is defined as $\pi D_{\rm
eff}/\lambda$. Similarly, for a dipole parallel to the plane, the
three partial contributions to the decay rate  can also be
expressed in terms of the parameter $\xi$ alone:
\begin{subequations}\label{separclosetoplane}
\begin{eqnarray}
\Gamma_{x}^{sr}(z_{\alpha},\Omega) & = &
\frac{3}{4}\Gamma_{0}\left[ 1 - \xi \arctan(1/\xi)\right]  \\
\Gamma_{x}^{pr}(z_{\alpha},\Omega) & = & \frac{3}{4
\xi^{2}}\Gamma_{0}\left[ 1-\frac{\arctan(\xi)}{\xi}\right]  \\
\Gamma_{x}^{sg}(z_{\alpha},\Omega) & = & \frac{3 \pi
\xi}{4}\Gamma_{0}.
\end{eqnarray}
\end{subequations} In Fig.~\ref{seratesxi} the relative rates are
plotted as a function of $\xi$.
The results can be checked in two limiting cases: if $D_{\rm
eff}=0$ there is no plane and then indeed both $\Gamma_{z}$ and
$\Gamma_{x}$ are equal to the free-space value $\Gamma_{0}$. The
other  limit is that of a perfect mirror, when  $D_{\rm eff}$ (and
consequently $\xi$) is sent to infinity. This limit is not visible
in the figure, but the limiting values are
$\Gamma_{z}/\Gamma_{0}=2$ and $\Gamma_{x}/\Gamma_{0}=0$. These
values indeed agree with the well-known emission rates for atoms
near perfect mirrors \cite{Drexhage70,Haroche92,Milonni94}.
Emission rates into guided modes vanish in the perfect-mirror
limit; at $z=z_{\alpha}$ this is only the case by sending $D_{\rm
eff}$ to infinity before putting $z$ equal to $z_{\alpha}=0$. This
completes the discussion of emission rates near a single plane.

\begin{figure*}[t]
\begin{center}
{\includegraphics[width=60mm,height=50mm]{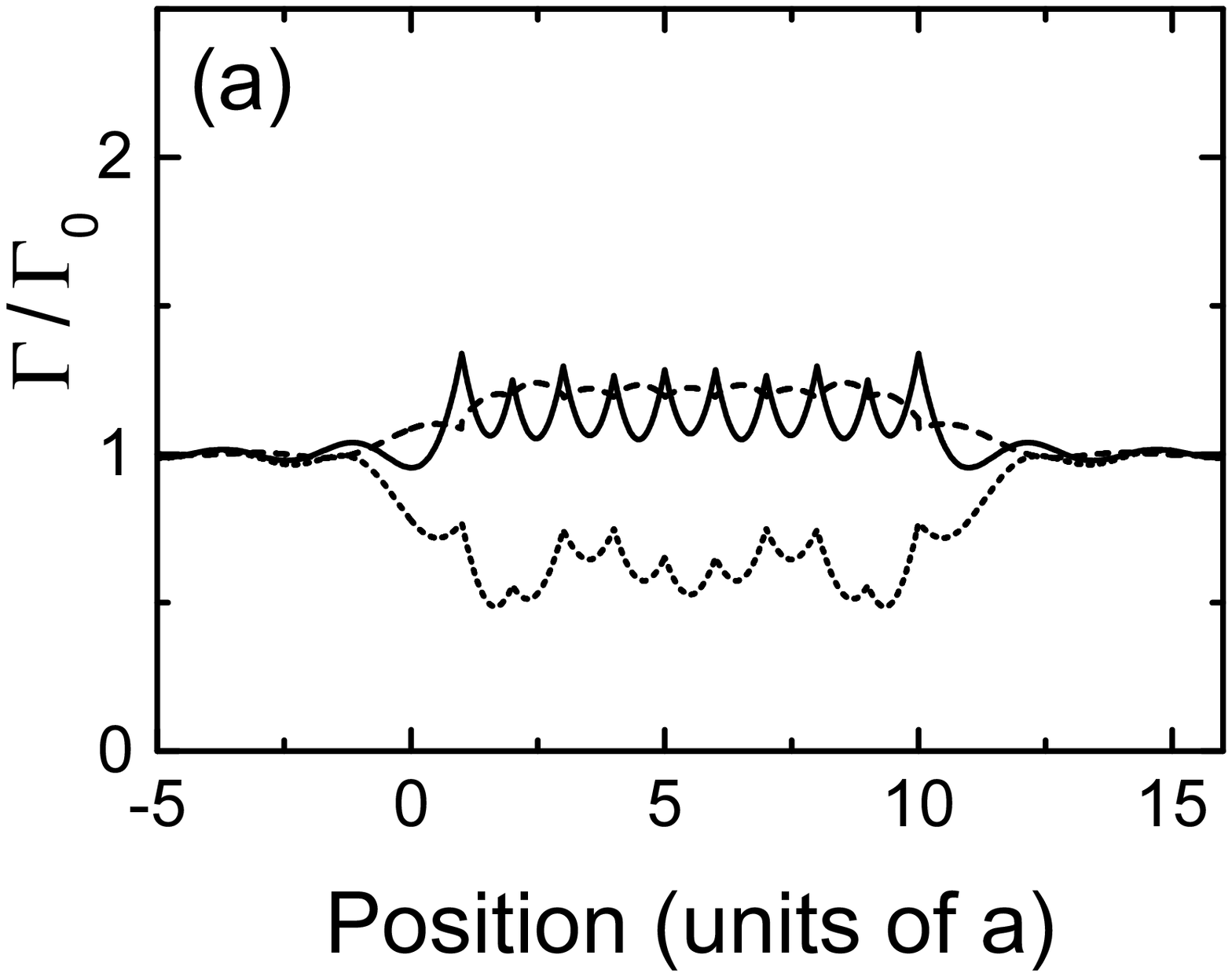}}
{\includegraphics[width=60mm,height=50mm]{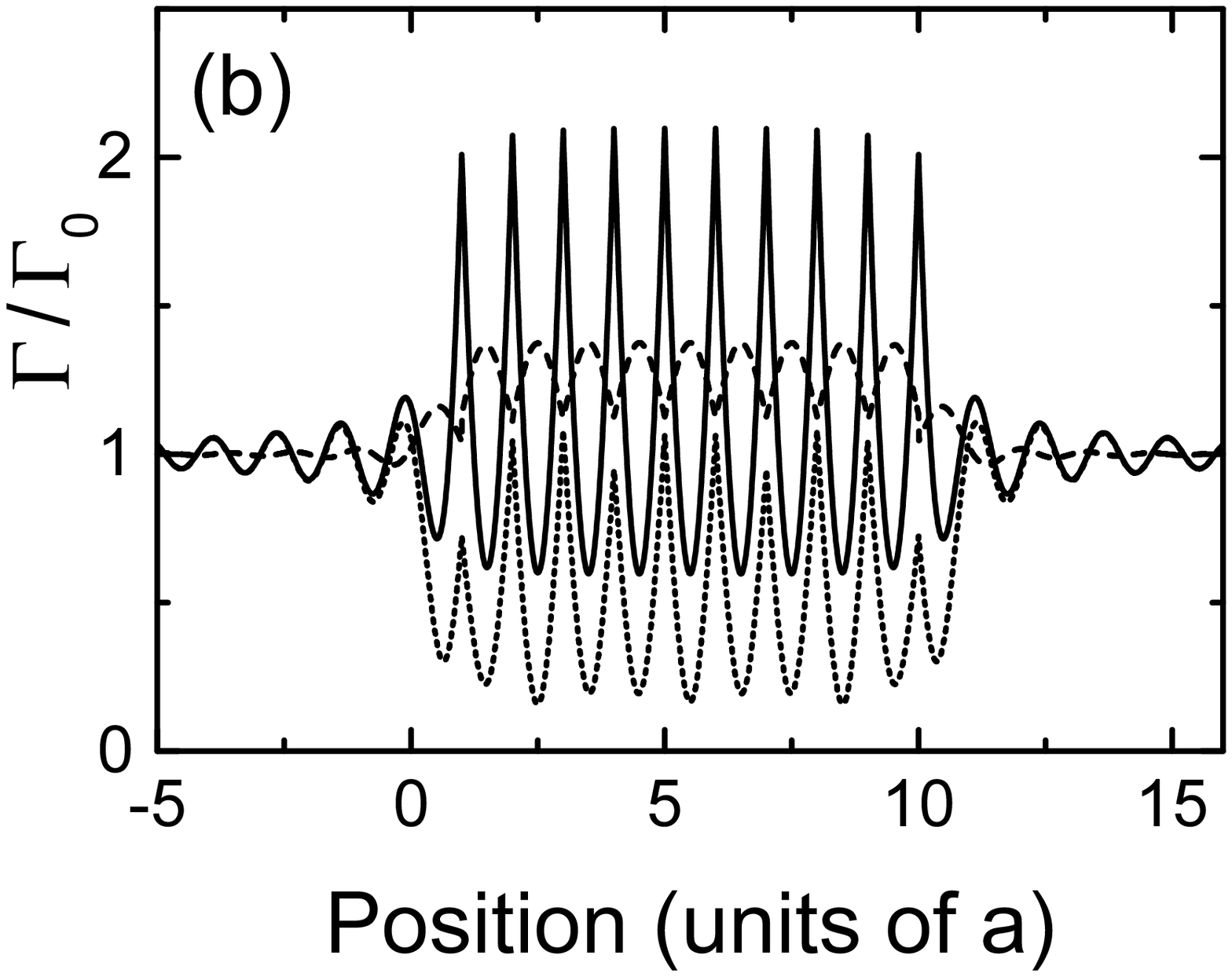}}
{\includegraphics[width=60mm,height=50mm]{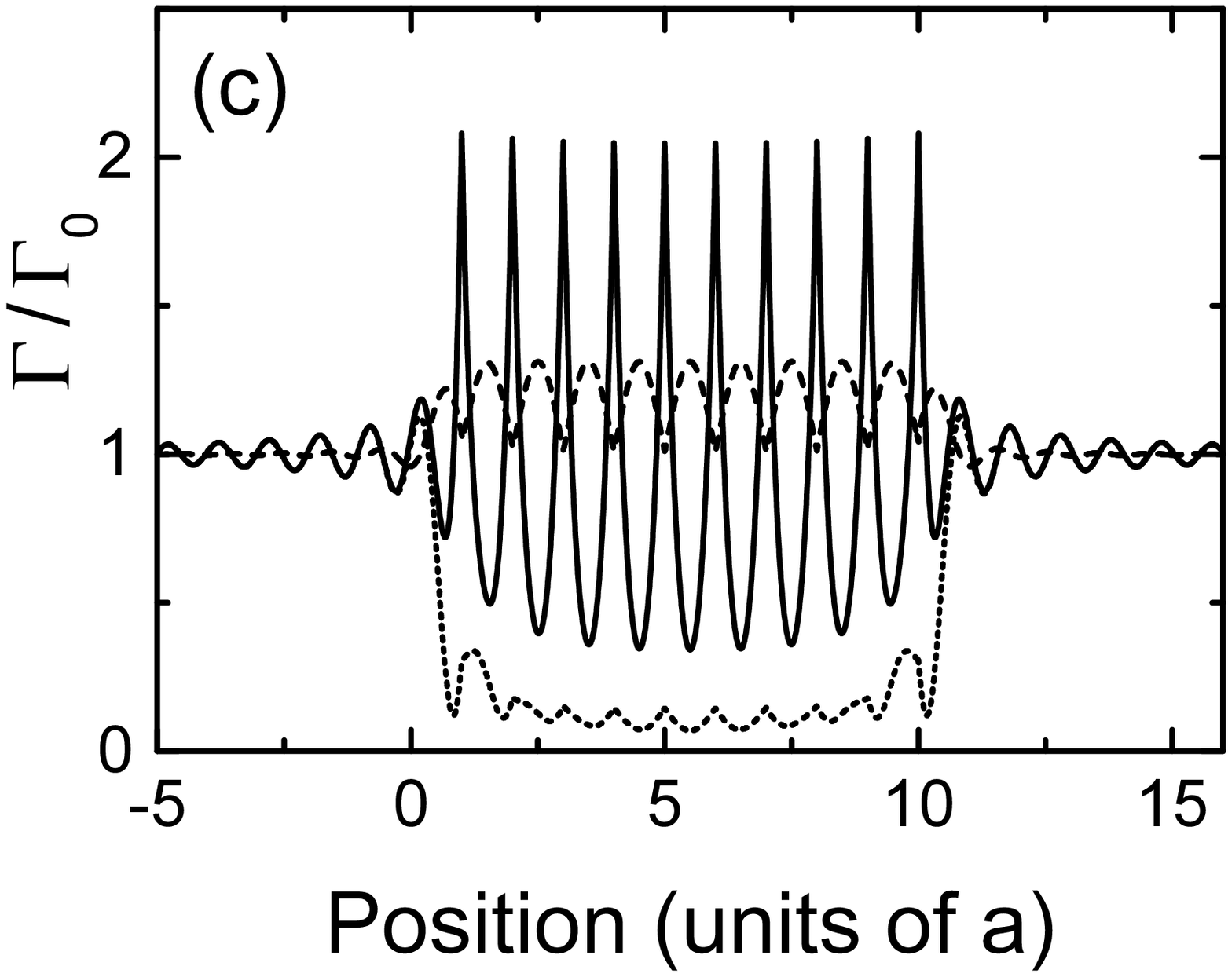}}
{\includegraphics[width=60mm,height=50mm]{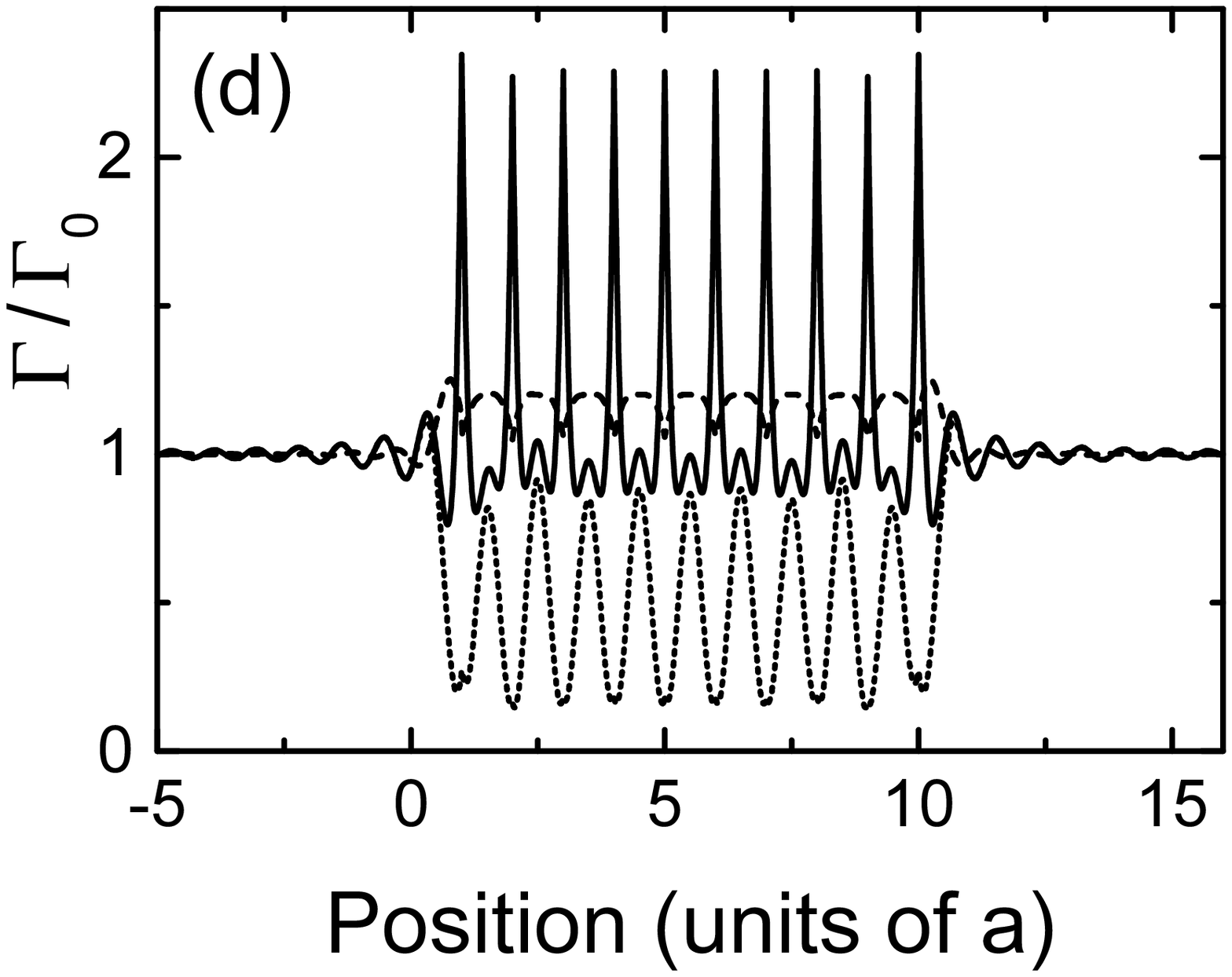}}
\end{center}
\caption{Spontaneous-emission rates $\Gamma_{x}$ (solid line) and
$\Gamma_{z}$ (dashed) near a ten-plane crystal. For all planes,
$D_{\rm eff} = 0.46 a$.  The dotted line is the radiative part of
$\Gamma_{x}$. The figures (a)-(d) correspond to four frequencies:
(a) $a/\lambda=0.2$; (b) $a/\lambda = 0.4$; (c) $a/\lambda=0.5$;
(d) $a/\lambda=0.6$.}\label{seratesN10}
\end{figure*}

Now consider emission rates inside and near  a crystal of a number
$N$ of plane scatterers. Results will be presented for $N=10$.  In
Fig.~\ref{seratesN10}(a-d), orientation-dependent
spontaneous-emission rates are plotted for several frequencies.
For clarity in the pictures, the positions of the planes at
$a,2a,...,10a$ are not shown as vertical lines this time. The most
striking difference between $\Gamma_{x}$ and $\Gamma_{z}$ is that
$\Gamma_{x}$ becomes very spiky near the planes, because only
parallel dipoles can couple to the $s$-polarized guided modes (and
because $p$-polarized guided modes are absent in our model).

As the frequency increases when going from
Fig.~\ref{seratesN10}(a) to Fig.~\ref{seratesN10}(d), $\Gamma_{x}$
becomes more spiky, because the partial emission rate into the
guided modes (the difference between the solid and the dotted
lines in Fig.~\ref{seratesN10}) becomes more concentrated near the
planes. The term `concentration' is appropriate here, because the
maximum amplitudes near the planes are higher for narrower spikes.
The same effect was observed for the single plane in
Fig.~\ref{serates05N1}(a-b), where the frequency is kept constant
and $D_{\rm eff}$ is increased instead.

The purely radiative-mode rate $\Gamma_{z}$ on average increases
due to  the presence of the planes, whereas the radiative part of
$\Gamma_{x}$ on average decreases. The same behavior occurs near a
single plane in Fig.~\ref{serates05N1} and for a perfect mirror.

Figure~\ref{modessquaredvp} showed that the optical modes of
$p$-polarized light have discontinuities at the plane positions.
There are also discontinuities in the spontaneous-emission rates
(not for a single plane, for symmetry reasons), but these are too
small to be visible in Fig.~\ref{seratesN10}. It can be understood
that they are small from the fact that the discontinuities per
mode are averaged in the emission rate.

 The dotted lines in Fig.~\ref{seratesN10} are the radiative parts of $\Gamma_{x}$.
These are similar to the radiative-mode rates for scalar waves
\cite{Wubs02}, but not identical, since in $\Gamma_{x}$ not only
$s$-polarized but also $p$-polarized light contributes, according
to Eq.~(\ref{gammax}). In particular, far away from the planes,
the emission rate of dipoles parallel to the planes consists of
$75\%$ $s$-polarized and $25\%$  $p$-polarized light. (To be sure,
 light emitted by perpendicular
dipoles is $100\%$ $p$-polarized for all layered dielectrics, see
Eq.~(\ref{gammaz}).)

 For `large enough'
photonic crystals, one expects inner unit cells to have optical
properties similar to unit cells in the infinite crystal. Are the
ten-plane crystals large enough? This  depends  on the properties
of a single plane. In two extreme cases, the  crystal size does
not matter: when the individual planes do not reflect any light in
any direction ($D_{\rm eff}/a=0$), or when they reflect all light
($D_{\rm eff}/a = \infty$), one finds the same emission rates in
finite and in infinite crystals, because for $D_{\rm eff}/a=0$ we
have the free-space case and for $D_{\rm eff}/a = \infty$ all unit
cells are optically disconnected. Only in intermediate cases
($0<D_{\rm eff}/a < \infty$) can finite photonic crystals have an
appreciable {\em and} unit-cell dependent influence on
spontaneous-emission rates.

In the intermediate cases (we assumed $D_{\rm eff}/a = 0.46$), the
single-plane reflection also depends on the frequency of the
light. The planes reflect light better at higher frequencies,
because material dispersion was neglected. It appears that for the
lowest frequency considered in Fig.~\ref{seratesN10},
$a/\lambda=0.2$ in (a), the emission rates in inner unit cells
already are influenced considerably by the crystal. The rates vary
in neighboring inner unit cells, which indicates that the rates
have not yet converged to the infinite-crystal values. For higher
frequencies, say $a/\lambda=0.6$, individual planes reflect light
much better. In the corresponding Fig.~\ref{seratesN10}(d), all
inner unit cells look alike and emission rates have converged.

Now consider a parallel dipole at a fixed position, very close to
a plane in an inner unit cell. From Fig.~\ref{seratesN10}, one can
also appreciate how the guided modes will influence the
(frequency-dependent) emission rate of a dipole very close to a
plane. For frequencies $a/\lambda=0.4$ and higher,
Figs.~\ref{seratesN10}(b-d) show that emission into propagating
modes modes is negligible compared to emission into guided modes.
 The dipole falls inside the guided-mode spike
near the plane. As the frequency increases, the maximum amplitude
of the spike increases. With the dipole still well inside the
spike, the emission rate of the dipole will increase as well. When
increasing the frequency further, the spike becomes so narrow that
the dipole ends up in one of the wings of the spike, until the
dipole finds itself completely outside it. This will cause
 the emission rates into guided modes to drop at higher frequencies. The combined
effect is a peak in the frequency-dependent emission rates.
Indeed, in \cite{Alvarado01} the dipole emission rate (or more
precisely, its $s$-wave component) for infinite crystals as a
function of frequency shows a pronounced peak for dipole positions
$z$ near a plane (see Fig.~3 in \cite{Alvarado01}). We can
unambiguously attribute this peak to emission into the guided
modes.

We can also understand how an emission  peak will depend on
$D_{\rm eff}$ and on the distance to the plane. We have seen that
spikes are narrower, with higher amplitudes, for larger $D_{\rm
eff}$ and for higher frequencies. When assuming larger $D_{\rm
eff}$, a dipole at fixed distance will feel a guided-mode enhanced
emission rate for lower frequencies. However, the dipole will also
at  lower frequencies begin to fall outside the range of the
spike. This explains why for fixed dipole position and increased
$D_{\rm eff}$, the emission peak has a larger amplitude and
attains its maximum at a lower frequency, precisely as seen in
Fig.~3 in \cite{Alvarado01}. Similar reasoning suggests that for a
parallel dipole a bit further away from a plane but still close to
it (while keeping $D_{\rm eff}$ fixed), an emission peak will
occur at lower frequencies, with lower amplitude.

What can be appreciated best in Fig.~\ref{seratesN10}(a) is that
for parallel dipoles the total emission rate converges faster than
the radiative and guided-mode partial rates separately. This can
be related to a kind of mode competition between $s$-polarized
propagating and guided modes that we also found for scalar waves
\cite{Wubs02}. On the other hand, for perpendicular dipoles all
emission is into radiative modes, so mode competition is absent
there and convergence sets in earlier.

For scalar waves  the ten-plane structure could act as an
omnidirectional mirror, whereas in Sec.~\ref{propmodesvector} it
was found that it is not an omnidirectional mirror for vector
waves. Correspondingly, the radiative-mode LDOS for scalar waves
at $a/\lambda=0.5$ dropped down to (almost) zero inside the
ten-plane omnidirectional mirror, whereas the emission rates in
Fig.~\ref{seratesN10}(c) show that the radiative LDOS for vector
waves stays nonzero inside the crystal. In the inner unit cells,
dipoles parallel to the planes emit predominantly guided light.
The small amount of light that leaves the structure is strongly
$p$-polarized. This is the case around $a/\lambda=0.5$ only, where
$s$-polarized light is omnidirectionally reflected. Such strongly
polarized emission is not a peculiarity of the plane-scatterer
model, because it will also occur for a real Bragg mirror whenever
light of only one of the two polarization directions is
omnidirectionally reflected. In the other plots in
Fig.~\ref{seratesN10} for higher and lower frequencies, the
radiative-mode parts of $\Gamma_{x}$ are the sums of emission
rates into both polarization directions.

\section{Conclusions, discussion, and outlook}\label{concchapvecplanes}
A  theory was set up for the multiple scattering of vector waves
by parallel planes, thereby  generalizing previous work for scalar
waves  to the more interesting but also more complicated case of
light waves. Unlike for scalar waves, the Green function had to be
regularized. This was accomplished by introducing a high-momentum
cutoff. An effective scattering theory emerged with a nonzero
T-matrix that no longer depends on the cutoff. The T-matrix and
Green-function formalism turned out to be very convenient for the
calculation of propagating and guided modes, as well as
spontaneous-emission rates, of finite photonic crystals of plane
scatterers.

A non-absorbing plane scatterer satisfies a separate optical
theorem for $s$- and $p$-polarized light. The radiative and guided
modes of $s$-polarized light could be mapped onto modes for scalar
waves. The $s$-polarized light has continuous modes, whereas
$p$-polarized modes  have discontinuities at the plane positions.

Throughout the paper, we have stressed the similarities and
differences of optical properties  of a plane scatterer or a
crystal of plane scatterers as compared to the corresponding
dielectric slab structures. It turns out that $p$-polarized waves
differ more in the two cases than $s$-polarized waves. Firstly,
because propagating $p$-polarized modes are different in the two
cases because the Brewster angle is $90^{\circ}$ for plane
scatterers. Secondly,  $p$-polarized guided modes in finite slab
structures have no analogues for plane scatterers. This was also
found in \cite{Shepherd97} for the single plane and in
\cite{Shepherd97,Zurita02} for the infinite crystal.

Unlike for scalar waves, equidistant and identical plane
scatterers can not be an omnidirectional mirror for all vector
waves. Such omnidirectional mirrors consisting of dielectric
layers do exist. For layered media, at least three refractive
indices are required in order to prevent complete transmission of
$p$-polarized light in the Brewster-angle direction
\cite{Fink98,Hooijer00,Lekner00}.

Omnidirectional reflection is a property of a dielectric for
external light sources. In this paper, omnidirectional reflection
could be related to the emission properties of atomic light
sources from within the finite crystals.  The graphs of emission
rates (Fig.~\ref{seratesN10}) show that the emission by dipoles
oriented parallel to the planes is affected much more strongly by
the planes than emission by perpendicular dipoles. This is a
characteristic of the plane-scatterer model, because the absence
of $p$-polarized guided modes is responsible for much of the
difference. In the frequency interval where $s$-polarized light is
omnidirectionally reflected, all light that exits the crystal
after a spontaneous-emission process will be $p$-polarized,
irrespective of the orientation of the emitters. Still, the major
fraction of the light will be emitted into guided modes and stay
inside the crystal.

For low frequencies, the single-plane reflectivity is lower,
emission rates are less affected by the crystal, and finite-size
effects are appreciable also in the inner unit cells of the
ten-plane crystal. For higher frequencies, planes reflect light
better and  emission rates are more strongly modified. In the
inner unit cells, the emission rates converge faster to the values
of the unit cell of an infinite crystal. If the infinite crystal
has larger variations in the emission rates inside a unit cell,
then a smaller finite crystal is needed to converge to this
result.

We argued that the guided modes will give rise to a peak in the
frequency-dependent emission rate of a parallel dipole close to a
 plane. We also reasoned that the peak will shrink and shift to
 lower frequencies when either the effective thickness or the
 distance to the plane is increased. Indeed, the occurrence of a
 peak and its dependence on the effective thickness agree with
 numerical calculations on infinite crystals in \cite{Alvarado01};
 some more numerical work is needed to
 corroborate our prediction of the distance dependence.

What other purposes can plane scatterers serve in the future? The
finite photonic crystals that can be made with them have
one-dimensional periodicity only, yet light propagation in all
three dimensions for all polarization directions is properly taken
into account. The number of planes can be chosen at will and
further advantages are that all optical modes and the complete
Green tensor can be determined.

Our calculations can be extended to situations where not al planes
are identical; or one could  allow light absorption or gain in the
planes by giving the effective thickness a complex value; the
number of planes per unit cell could also be increased to more
than one. Such calculations are possible in our formalism because
the results (\ref{Tpexactlysummed}) and (\ref{tNexactcomps})  for
the T-matrix were generalized in Eq.~(\ref{tnfinalapp}) of the
Appendix to planes chosen at arbitrary positions, each with a
different T-matrix $\bfsfT_{\alpha}({\bf k}_{\parallel}, \omega)$,
in other words with a different effective thickness. The model can
therefore also be used to study the effects of disorder in the
positions or in the optical properties of the planes on the
spontaneous-emission rates of embedded atoms.

 It would also be
interesting  to study models of finite photonic crystals built up
of non-parallel planes. This can be done, but numerically the
model would become more involved,   not so much because the planes
have an overlap of measure zero, but rather because the plane
representations for nonparallel planes will be different.
Numerical calculations for nonparallel planes require the
discretization of individual planes.

Apart from extending it, one can extract other interesting
observables from the model. For example, the knowledge of the
complete Green function makes it possible to calculate both
far-field and near-field spectra of  atoms embedded in the finite
crystal. It would be interesting to study spectra near frequencies
where the corresponding infinite crystal gives rise to a Van Hove
singularity in the emission rates \cite{Zurita02}. Crystals of
plane scatterers can serve as a model environment to study the
modification of several quantum optical processes when atoms are
embedded in photonic crystals. Transient effects in the
spontaneous-emission rates are but one  example, thereby
generalizing work done on a one-dimensional cavity formed by two
planes \cite{Jedrkiewicz99}. Calculations are underway that show
photonic-crystal induced modifications of cooperative atomic
processes.

\section*{Acknowledgements}
We would like to thank Rudolf Sprik and Willem Vos for stimulating
discussions. This work is part of the research program of the
Stichting voor Fundamenteel Onderzoek der Materie, which is
financially supported by the Nederlandse Organisatie voor
Wetenschappelijk Onderzoek.

\appendix
\section*{Appendix: Derivation of N-plane T-matrix} The goal of this appendix
is to derive the
expressions~(\ref{Tpexactlysummed}-\ref{tNexactcomps}) for the
$N$-plane T-matrix $T^{(N)}$, by summing the infinite series in
Eq.~(\ref{Tgeneralintsingles}), where the $\bfsfT_{\alpha}$ are
the T-matrices of single planes:
\begin{equation}\label{Tplaneapp}
\bfsfT_{\alpha}(\omega)= \frac{1}{(2\pi)^{2}}\int\mbox{d}^{2}{\bf
k}_{\parallel}\; |{\bf k}_{\parallel}, z_{\alpha}\rangle
\bfsfT_{\alpha}({\bf k}_{\parallel},\omega)\langle z_{\alpha},{\bf
k}_{\parallel}|.
\end{equation}
The $\bfsfT_{\alpha}({\bf k}_{\parallel},\omega)$ are  $3\times 3$
tensors, defined in Eq.~(\ref{Eeffectivedescript}). First allow
all planes to have  different $\bfsfT_{\alpha}({\bf
k}_{\parallel},\omega)$; furthermore, allow the parallel planes to
have arbitrary (but all different) positions $z_{\alpha}$.

The first-order term in the expansion~(\ref{Tgeneralintsingles})
is simply the sum of the $\bfsfT_{\alpha}$ of
Eq.~(\ref{Tplaneapp}); the second-order term has the form
\begin{equation}\label{secondorderexpapp}
\frac{1}{(2\pi)^{2}}\int\mbox{d}{\bf k}_{\parallel}
\sum_{\alpha,\beta=1}^{N}|{\bf k}_{\parallel}, z_{\alpha}\rangle
\bfsfT_{\alpha}({\bf k}_{\parallel};\omega)\cdot\left[\bfsfD({\bf
k}_{\parallel}, \omega)\right]_{\alpha\beta} \langle {\bf
k}_{\parallel},z_{\beta}|,
\end{equation}
where the dot denotes the inner product of $3\times 3$ tensors.
Furthermore, the property $\langle {\bf k}_{\parallel}, z|
\bfsfG_{0}(\omega)|{\bf k'}_{\parallel},z'\rangle =
(2\pi)^{2}\delta^{2}({\bf k}_{\parallel}-{\bf k'}_{\parallel})
\bfsfG_{0}({\bf k}_{\parallel}, z, z',\omega)$ of the free-space
Green function was used. The  matrix elements
$\bfsfD_{\alpha\beta}$ of the matrix $\bfsfD$ are $3\times 3$
tensors which are defined as
 \begin{equation}\label{Ddefapp}
 \left[\bfsfD({\bf k}_{\parallel}, \omega)\right]_{\alpha\beta} \equiv
(1-\delta_{\alpha\beta})\bfsfG_{0}({\bf k}_{\parallel},
z_{\alpha},z_{\beta},\omega)\cdot \bfsfT_{\beta}({\bf
k}_{\parallel}, \omega).
\end{equation}
The third-order term of the expansion~(\ref{Tgeneralintsingles})
is as Eq.~(\ref{secondorderexpapp}), with $\bfsfD_{\alpha\beta}$
replaced by $\bfsfD_{\alpha\beta}^{2}=\sum_{\gamma=1}^{N}
\bfsfD_{\alpha\gamma}\cdot\bfsfD_{\gamma\beta}$, that is, the
square of the $N \times N$ matrix $\bfsfD$. Similarly, the
fourth-order term features the cube of $\bfsfD$, and so on. By
summing all orders, we have
\begin{equation}\label{Tpexactlysummedapp}
\bfsfT^{(N)}(\omega) = \frac{1}{(2\pi)^{2}}\int\mbox{d}^{2} {\bf
k}_{\parallel}\sum_{\alpha,\beta=1}^{N}|{\bf
k}_{\parallel},z_{\alpha}\rangle \bfsfT_{\alpha\beta}^{(N)}({\bf
k}_{\parallel},\omega) \langle {\bf k}_{\parallel},z_{\beta}|,
\end{equation}
where the $N$-plane T-matrix $\bfsfT_{\alpha\beta}^{(N)}({\bf
k}_{\parallel},\omega)$ is given by the series
\begin{eqnarray}
\bfsfT_{\alpha\beta}^{(N)}({\bf k}_{\parallel},\omega)&  = &
\bfsfT_{\alpha}\cdot\left[ \bfsfI \otimes \bfsfI_{N} + \bfsfD +
\bfsfD^{2} +
\bfsfD^{3} +\ldots\right]_{\alpha\beta} \nonumber  \\
& = & \bfsfT_{\alpha}\cdot\left[ \bfsfI \otimes \bfsfI_{N} -
\bfsfD \right]_{\alpha\beta}^{-1}.
\end{eqnarray}
Here, $\bfsfI_{N}$ is the $N\times N$ unit matrix and, as before,
$\bfsfI$ is the $3 \times 3$ unit tensor, so that $\left[\bfsfI
\otimes
\bfsfI_{N}\right]_{\alpha\beta}=\delta_{\alpha\beta}\bfsfI$. Now
in the plane representation,  the Green tensor has the form as
derived in Sec.~\ref{greenoplane},  and the single-plane
T-matrices $\bfsfT_{\alpha}({\bf k}_{\parallel},\omega)$ have the
simple form as defined in Eq.~(\ref{Eeffectivedescript}), with
only $T_{\alpha}^{ss}$ and $T_{\alpha}^{vv}$ nonzero. It follows
that $\bfsfD_{\alpha\beta}$,  $\bfsfD_{\alpha\beta}^{2}$, and all
higher powers of $\bfsfD$ have only three nonzero spatial
components, namely the $ss$-, $vv$-, and the $zv$-components. As a
consequence, $\bfsfT_{\alpha\beta}^{(N)}({\bf
k}_{\parallel},\omega)$ has only two nonzero spatial components
for all plane indices $\alpha$ and $\beta$, namely the $ss$- and
$vv$-component:
\begin{equation}\label{tnfinalapp}
\bfsfT_{\alpha\beta}^{(N)}({\bf k}_{\parallel},\omega)  =
\left(\begin{array}{ccc}
T_{\alpha}^{ss}\left[\bfsfI_{N}-\bfsfD^{ss}\right]_{\alpha\beta}^{-1} & 0 & 0 \\
0 & T_{\alpha}^{vv}\left[\bfsfI_{N}-\bfsfD^{vv}\right]_{\alpha\beta}^{-1} & 0 \\
0 & 0 & 0
\end{array}\right).
\end{equation}
This generalizes the result of Eq.~(\ref{Eeffectivedescript}) for
the single-plane T-matrices $\bfsfT_{\alpha}$, where the same two
 components were found to be nonzero. The generalization is from one
plane to $N$ parallel planes placed at arbitrary positions, each
plane possibly with a different effective thickness. In the
special case considered in the main text that all planes are
identical, in other words if all $\bfsfT_{\alpha}({\bf
k}_{\parallel},\omega)$ are equal, then Eq.~(\ref{tnfinalapp})
simplifies into Eq.~(\ref{tNexactcomps}).

\end{document}